\documentclass[pdflatex,sn-nature]{sn-jnl}

\usepackage{graphicx}%
\usepackage{amsmath,amssymb,amsfonts}%
\usepackage{booktabs}%
\usepackage{tikz}
\usepackage{bm}
\usepackage[dvipsnames]{xcolor}
\usepackage[symbol]{footmisc}

\graphicspath{{figures/}}

\begin{document}


\title[Direct numerical simulations of inhalation in a 23-generation lung model]{Direct numerical simulations of inhalation in a 23-generation lung model}

\author*[1]{\fnm{Marco} \sur{Atzori}}\email{marco.atzori@polimi.it}
\author[1]{\fnm{Emanuele} \sur{Gallorini$^\dagger$}} 
\author[2]{\fnm{Ciro} \sur{Cottini}} 
\author[2,3]{\fnm{Andrea} \sur{Benassi}}
\author[1]{\fnm{Maurizio} \sur{Quadrio}}

\affil*[1]{\orgdiv{Department of Aerospace Science and Technology}, \orgname{Politecnico di Milano}, \orgaddress{\street{Via La Masa, 34}, \city{Milano}, \postcode{20156}, \country{Italy}}}

\affil[2]{\orgname{Chiesi Farmaceutici S.p.A.}, \orgaddress{\street{Largo Belloli 11A}, \city{Parma}, \postcode{43122}, \country{Italy}}}

\affil[3]{\orgname{International School for Advanced Studies (SISSA)}, \orgaddress{\street{via Bonomea, 265}, \city{Trieste}, \postcode{34136}, \country{Italy}}}


\addtocounter{footnote}{2}%
\footnotetext{Current address: Univ. Lille, CNRS, ONERA, Arts et M\'etiers ParisTech, Centrale Lille, UMR 9014 – LMFL – Laboratoire de M\'ecanique des Fluides de Lille – Kamp\'e de Feriet, F-59000 Lille, France}

\abstract{
The air flows in the proximal and distal portions of the human lungs are interconnected: the lower Reynolds number in the deeper generations causes a progressive flow regularization, but mass conservation requires flow rate oscillations to propagate through the airway bifurcations.
To explain how these competing effects shape the flow state in the deeper generations, we have performed the first high-fidelity numerical simulation of the air flow in a lung model that includes 23 successive bifurcations of a single planar airway. Turbulence modelling or assumptions on flow regimes are not required.
The chosen flow rate is steady on average, and representative of the peak inspiratory flow reached by adult patients breathing through therapeutical inhalers.  
As expected, advection becomes progressively less important after each bifurcation, until a time-dependent Stokes regime governed solely by viscous diffusion is established in the smallest generations. 
However, fluctuations in this regime are relatively fast and large with respect to the mean flow, which is in contrast with the commonly agreed picture that only the breathing frequency is relevant at the scale of the alveoli.
We demonstrate that the characteristic frequency and amplitude of these fluctuations are linked to the flow in the upper part of the bronchial tree, as they originate from the time-dependent flow splitting in the upper bifurcations. 
Even though these fluctuations are observed here in an idealized, rigid lung model, our findings suggest that the assumptions usually adopted in many of the current lung models might need to be revised. 
}


\keywords{Respiratory Flows, Lung models, Deep airways, Computational Fluid Dynamics (CFD), Direct Numerical Simulations (DNS)}

\maketitle



%




\section{Introduction}
\label{sec:intro}
The lungs are traditionally divided into a purely conducting upper zone and a deeper functional region, progressively populated by alveoli, where gas exchange between air and blood takes place. 
The trachea branches first into the main bronchi, followed by the lobar and segmental bronchi, and by progressively smaller branches, until the first alveolated ducts appear in the acini. The airways are terminated by capping alveolar sacs, after an average of 22--23 subsequent bifurcations which originate a large number of individual ducts, known as generations, and a complex, self-affine structure~\cite{weibel-gomez-1962, pedley-1977, finlay-2019}.
An in-depth comprehension of the fluid mechanics of this system is crucial to understand the physiology of the lungs, to mitigate risks associated with inhalation of pathogens and pollutants~\cite{neelakantan-etal-2022}, and to effectively develop therapeutic aerosols~\cite{kole-etal-2023}. 
Unfortunately, such complexity poses formidable challenges to both experimental measurements and numerical simulations. 
Considering as an example aerosol inhalation, we can currently predict deposition in the deep airways by using computational fluid dynamics~\cite{koullapis-etal-2019}, statistical models~\cite{yeh-schum-1980, koblinger-hofmann-1990} or hybrid approaches \cite{grill-etal-2023,zhang-etal-2022} that require tuning of a large number of parameters; lack of high-quality data remains a pressing issue that hinders both calibration of the existing models and development of better ones~\cite{walenga-etal-2023}.

High-fidelity, model-free numerical simulations of the air flow down to the lowest lung generations have been so far unattainable. It is unclear whether and to what extent the chaotic flow in the upmost portion of the bronchial tree influences the smallest ducts in the lowest generations.
In this work, we employ a 23-generations lung model similar to that experimentally considered by M\"oller et al.~\cite{moller-etal-2021}, and thanks to a newly developed, efficient numerical solver for the incompressible Navier--Stokes equations we carry out a direct numerical simulation (DNS) at an average flow rate of $\approx 60$ L/min, a value representative of the inspiratory peak for an adult patient breathing through an inhaler for therapeutic aerosol delivery. 
For the first time, an idealized branch from the scale of the trachea to that of the alveoli becomes observable in its entirety, with a fully time- and space-resolved simulation free of modelling assumptions. 

We give an overview of relevant previous findings on fluid mechanics of the bronchial tree in \S\ref{sec:intro_flow} and on the approaches adopted by the most recent numerical studies addressing deep generations in \S\ref{sec:intro_num}; our numerical method and case study are described in \S\ref{sec:methods}; we present our results in \S\ref{sec:results}; finally, we summarize our findings in the concluding discussion \S\ref{sec:conclusions}.

\subsection{Flow regimes in the human lungs}
\label{sec:intro_flow}
Significant anatomical variability is observed in the lungs, for instance between sexes~\cite{christou-etal-2021}, but also among subjects otherwise similar in physical constitution~\cite{kassinos-sznitman-2024}. 
The most prominent feature is the large number of bifurcations through which the airways morph from the trachea to the alveolar region. Shapes and diameters across generations are similar enough among subjects~\cite{haefeli-bleuer-weibel-1988}, so that creating generic lung models is recognized as a reasonable approach since a long time~\cite{heistracher-hofmann-1995,schmidt-etal-2004}.
Therefore, the fluid mechanics of such models is interesting in general terms.  

Thanks to the relatively low air speed, the flow is incompressible~\cite{pedley-1977}. 
Moreover, in breathing under standard conditions, nitrogen does not participate to the gas exchange with the blood, and variations of temperature and humidity, although relevant from a physiological perspective, are not large enough to affect the mechanical properties of the air. 
Air density $\rho$ and kinematic viscosity $\nu$ can be thus considered constants~\cite{kleinstreuer-zhang-2010}. 
It follows that the governing equation for the air flow are the incompressible Navier--Stokes equations written for a constant-properties fluid. One relevant dimensionless group is the Reynolds number, typically defined as $Re=\mathcal{U} \mathcal{L} /\nu$ (where $\mathcal{U}$ and $\mathcal{L}$ are characteristic velocity and length scales). 
The approximately periodic nature of breathing also suggests the use of the Womersley number, defined as $Wo = \mathcal{L} \sqrt{2 \pi \mathcal{F} / \nu}$, where $\mathcal{F}$ is a characteristic frequency. 

The flow regime in the lungs can be hinted at via the characteristic values of $Re$ and $Wo$ in each lung generation~\cite{pedley-1977}.
In the trachea, $Re$ reaches up and above $\approx 5000$, which is high enough for turbulence to exist. Moreover, the Reynolds number inside inhalers is of the order of $Re\approx O(10^5)$, where significant pressure and velocity fluctuations are generated on purpose to achieve good quality aerosols, and then propagate towards the airways. 
For several successive bifurcations, $Re$ remains large enough for the existence of chaotic flow states, as suggested by studies on simplified geometries including both single~\cite{jan-shapiro-kamm-1989} and multiple~\cite{suh-park-2018} bifurcations. 
This is confirmed by experiments~\cite{banko-etal-2016} and high-fidelity simulations~\cite{bernate-etal-2017} on more realistic models, often down to generations 5 or 8, for both constant inhalation and sinusoidal breathing. 
To reduce computational cost, numerical studies in most cases employ the so-called large-eddy simulation (LES) approach, where the simulation is time-resolved but turbulence models are used to account for the small (spatial and temporal) turbulent scales that are not captured by the numerical discretization.
With LES, Jing et al.~\cite{jing-etal-2023} have recently reached down to generation $13$.
Several works, limited to the upper portion of the bronchial tree, have also considered particles dispersion, as in the numerical and experimental campaigns of the SimInhale benchmark~\cite{koullapis-etal-2018, janke-etal-2019, spasov-etal-2024}, designed to provide reference data for lower-fidelity methods. 

In the lower generations, the airways becomes smaller, and the value of $Re$ diminishes below what is required to generate turbulent fluctuations. 
If the breathing time scale is assumed to be the dominant frequency across all generations, $Wo$ decreases in a similar way.
Eventually, both $Re$ and $Wo$ become so small that a chaotic regime cannot be sustained, and the flow becomes laminar~\cite{sznitman-2013}.
If ducts are regular enough, the time dependency of the velocity manifests simply as a rescaling of the laminar solution, following the breathing cycle~\cite{haber-etal-2000}.
Starting (on average) at generation $14$, alveoli begin to populate the airway surfaces, appearing after every bifurcation with increasing density until the terminal alveolar sacs are eventually reached~\cite{weibel-sapoval-filoche-2005}, and the geometries typical of the alveolar regions lead to more complex flow patterns, including time-dependent separation~\cite{shen-etal-2020}.

\subsection{Simulations including deep generations}
\label{sec:intro_num}
Extending the numerical approaches used to compute the flow in the upper generations down to the lowest generations has proved so far to be unfeasible. 
Scale separation, on the one hand, with duct diameters spanning between $\approx$ $2 \, \rm cm$ and $\approx 200 \,\mu \rm m$ across the lungs, and the number of branches, on the other hand, lead to prohibitive computational costs.
The prevailing approach, therefore, is to start from a realistic description of the flow down to an intermediate generation, and describe lower generations with stochastic models, to obtain e.g. estimates of particle deposition~\cite{yeh-schum-1980, koblinger-hofmann-1990, goo-kim-2003, sonnenberg-etal-2020} or ventilation distribution~\cite{swan-clark-tawhai-2012}. 
Such models, however, do not provide information regarding the transition between flow regimes in the bronchial tree, for which solving the governing equations is unavoidable. 

Alternatively, one could try and reach the terminal generations with numerical approaches of lesser computational cost, which include among others: following a limited number of branches; using turbulence models more aggressively to reduce resolution requirements; and using different methods in different portions of the bronchial tree. 
Tian et al.~\cite{tian-etal-2011,longest-etal-2016} developed the so-called stochastic individual path (SIP) model, where a realistic geometry including the throat and a single branch down to generation $15$ is coupled with a simplified model for the acinar region below. Unsteady Reynolds-Averaged Navier--Stokes (URANS) simulations were used for the portion of bronchial tree down to generation $3$, providing time-averaged velocity profiles used as inlet condition for steady simulations in deeper generations.
URANS simulations differ from LES: both model turbulence to some extent, but the modelling extent is larger in URANS, further reducing computational costs at the price of lower fidelity~\cite{pope-2000}.  
Koullapis et al.~\cite{koullapis-etal-2018a} carried out LES with particle deposition in generations from $10$ to $19$, using a laminar pulsatile flow as inlet at generation $10$, with a peak flow rate of $25.6 \, \rm L/min$, and a model of idealized acinar region at each outlet of generation $19$. 
Islam et al.~\cite{islam-etal-2017} also carried out LES in generations from $0$ to $17$, with a steady laminar inlet at generation $0$ (a section of the trachea) and flow rates up to $60 \, \rm L/min$.  
Kolanjiyil and Kleinstreuer~\cite{kolanjiyil-kleinstreuer-2016} employed a model extending from the oral cavity to generation $23$ for two branches, and carried out URANS simulations, with flow rates of $15 \, \rm L/min$; the RANS turbulence model was kept active down to generation $6$.

Such approaches are particularly appealing for studying particle dispersion and deposition, as only particles smaller than a certain size are sensitive to small-scale fluctuations~\cite{koullapis-etal-2018}. 
In general, however, the flow in a large portion of the domain tends to remain (or it is just assumed to be) laminar, and the only frequency that is allowed to appear in the deep generations is the breathing frequency itself. 
This is reasonable, as long as a laminar inflow is used; whether it is also realistic depends on how the transition between different flow regimes actually takes place in the lungs. Of particular concern is the artificial regularization introduced by the turbulence model, that might alter the propagation of disturbances.

\section{Methodology}
\label{sec:methods}

\subsection{Test case}

\begin{figure*}
\centering
\includegraphics[width=.3\textwidth]{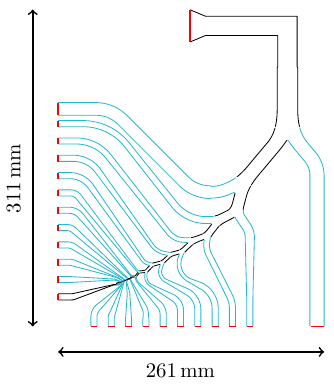}
\includegraphics[width=.66\textwidth]{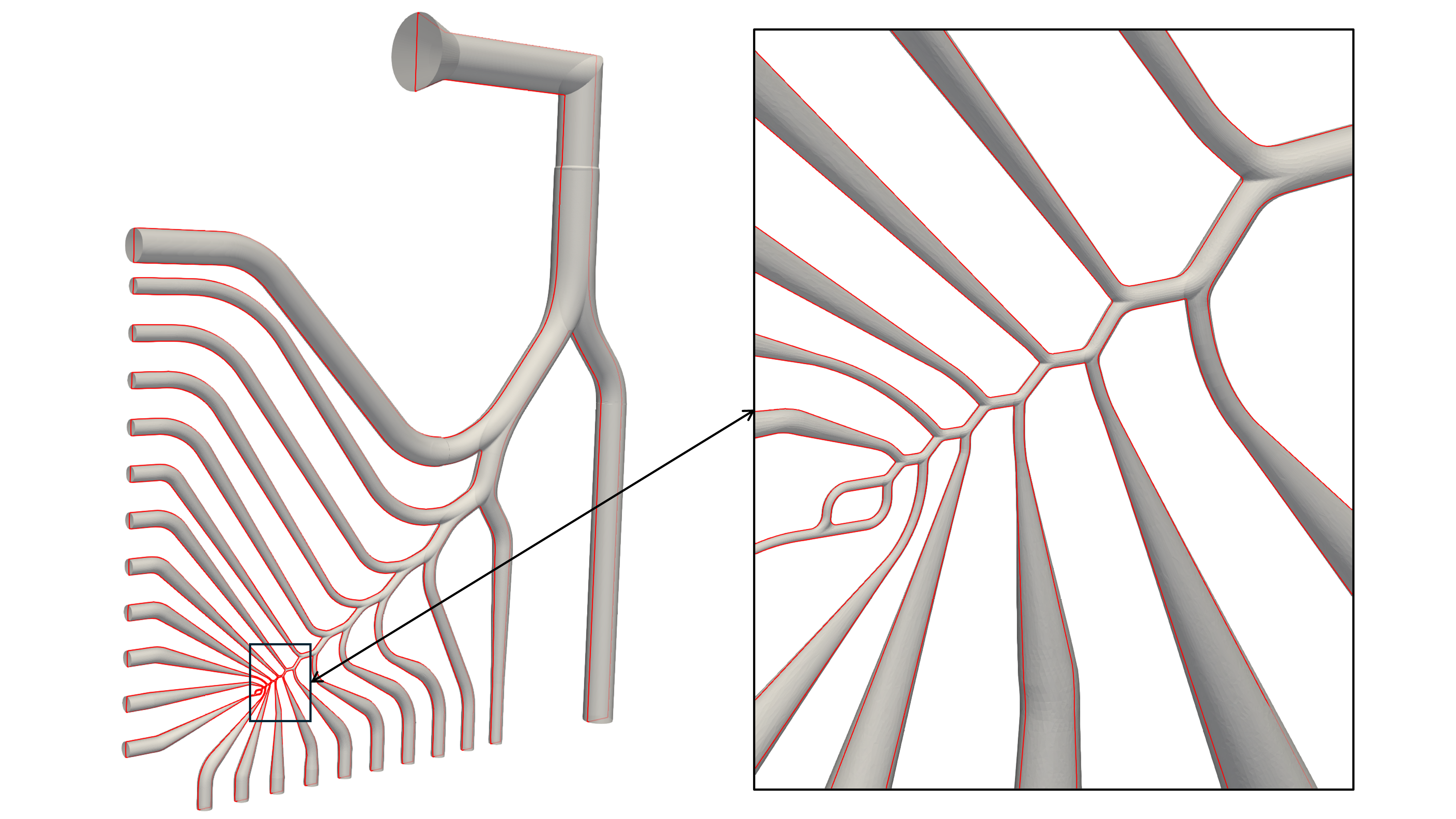}
\caption{The lung model adopted in the present work, derived from Ref.~\cite{moller-etal-2021}: frontal section at the symmetry plane (left), three-dimensional view (center), and zoom on the deepest generations (right). In the left panel, black and cyan denote the main branch and the secondary branches, respectively; the inlet and outlet patches are drawn in red. The symmetry plane shown in the left panel is highlighted in the views in the center and right panels by the red lines.}
\label{fig:model_overview}
\end{figure*}

The geometry employed in this study is the one created by M\"oller et al.~\cite{moller-etal-2021} for their experimental campaign, and consists of a single branch that bifurcates $23$ times. At every bifurcation, the flow rate is supposed to split approximately in half between the primary branch at the lower generation and a secondary branch that goes directly to the outlet, as illustrated in Fig.~\ref{fig:model_overview} (left).
The model is rigid and planar, i.e.\ the centre line of every branch lies on the sole symmetry plane of the model.
Two minor changes have been introduced to the model. The upper part of the airways is upgraded to reproduce the so-called United States Pharmacopoeia Induction Port (USP-IP), a simple geometry, made by a truncated cone followed by two perpendicular straight pipes~\cite{USPGeneral2012}, that induces the sudden change of direction experienced by the airflow in crossing the trachea. 
Moreover, the terminal portion of the two highest secondary branches is enlarged, so that the local velocity remains limited, with benefits in terms of computational cost.

For the remainder of this document, we will refer to the generation number with the integer $g$, where $g=0$ identifies the branch before the first bifurcation; furthermore, subscripts are reserved for generation numbers of generation-specific physical quantities.
Hence, for example the volumetric flow rate $q$ through the inlet of the main branch is denoted as $q_0$, and the flow rate through a generic generation of the main branch as $q_g$. 
The flow rates through the generic secondary branch is denoted as $q^s_g$.
Owing to mass conservation, the sum of the flow rates through each generation of the main branch and the corresponding secondary branch must equal the flow rate in the previous generation before splitting, i.e.\: $q_g=q_{g+1}+q^s_{g+1}$. 
In general, $q=q(t)$ is a function of time, for which a time-averaged value $Q={\rm Mean}[q]$, and fluctuations w.r.t.\ the average, $q'(t)=q(t)-Q$ can be defined, in the spirit of the so-called Reynolds decomposition of turbulent variables.
This work considers $Q_0 = 10^{-3} \, \rm m^3/s$, or $60 \, \rm L/min$, and aims at a flow splitting $Q_{g+1}/Q_g = 0.5$ for all bifurcations. 

\subsection{Numerical method}
The governing equations are the incompressible Navier--Stokes equations, expressing mass conservation and momentum balance:
\begin{equation}
\begin{split}
\bm{\nabla} \cdot \bm{u} &=0\,,\\
\frac{\partial \bm{u}}{\partial t} + \left( \bm{u} \cdot \bm{\nabla} \right) \bm{u} &= - \frac{1}{\rho} \bm{\nabla} p + \nu \nabla^2 \bm{u}
\end{split}
\label{eqn:gov}
\end{equation}
where $\bm{u}$ and $p$ denote the instantaneous velocity vector (with magnitude $u$) and pressure, and $\rho$ is the fluid density. 
The equations above are numerically solved without turbulence modelling, by resorting to spatial and temporal resolution fine enough to capture all the dynamically significant scales of turbulence: this is the most computationally demanding approach, known as direct numerical simulation (DNS). 
The solver is a new high-performance numerical code based on the immersed-boundary method (IBM), where the computational grid does not need to conform to the boundary. 
It is written in the CPL programming language~\cite{luchini-2021}, and was described in the recent Ref.~\cite{luchini-etal-2025}. 
The governing equations~\eqref{eqn:gov} are discretized on a staggered grid, using the centred second-order approximation for the spatial derivatives. 
Time integration of the momentum equation is performed using a third-order Runge--Kutta scheme, implemented to allow either for a constant time step or a constant Courant--Friedrichs--Lewy (CFL) number, and the pressure gradient is updated pointwise to fulfil the continuity constraint.
The streamlined implementation of the IBM leads to a notable efficiency of the code: the computational cost is low at about $10^{-7}\,\rm s$ of wall-clock time per time step and grid point. 

\subsection{Setup of the simulations}
Three crucial aspects of the computational procedures warrant a detailed description: a) the boundary conditions; b) the design of the computational grid; and c) the statistical significance of the results. 

\subsubsection{Boundary conditions}
\begin{figure}[htp]
\centering
\includegraphics[width=0.5\textwidth]{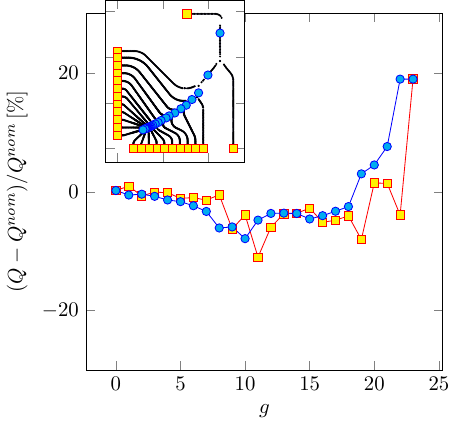} 
\caption{Relative difference of averaged flow rates w.r.t.\ the nominal values for the ideal $50/50$ splitting. As illustrated in the insert, yellow squares and blue dots indicate measurements at inlet and outlet patches or sections of the main branch, respectively.}
\label{fig:flow_rates}
\end{figure}

Once $Q_0 = 10^{-3}\, \rm m^3/s$ is set, the range of nominal flow rates through the descending generations, denoted with $Q^{nom}_g$, is computed dividing by powers of $2$.
Hence, $Q^{nom}_g = Q_0/2^g$ spans from $5 \cdot 10^{-4}\, \rm m^3/s$ for $g=1$ to $2.38 \cdot 10^{-10}\, \rm m^3/s$ for $g=23$. 
High precision in controlling mean flow rates is therefore necessary, since a $0.01\%$ inaccuracy in $Q_0$ corresponds to $100\%$ of $Q_{23}$.
Propagation of disturbances across the domain, especially in an incompressible simulation, makes the choice of boundary conditions a delicate task. 
To minimize artificial fluctuations, we employ Neumann-type boundary conditions for the velocity and constant pressure at inlet and outlets. Mimicking the actual experiment \cite{moller-etal-2021}, where the outflow from each secondary branch is regulated by a valve, the distribution of flow rates is driven by the pressure at the outlets, whereas the inlet reference pressure is set to zero.
The required outlet pressures are determined with a preliminary set of simulations where the pressure values at the boundary conditions were calibrated to achieve a $\approx50/50$ flow splitting at each bifurcation. 
Flow rates through the main and secondary branches have been checked {\em a posteriori} against the nominal values, using measurements both at inlet and outlet patches as well as at a section of each generation, as shown in Fig.~\ref{fig:flow_rates}.
Note that $Q$ measured at the inlet patch and the $23^{\rm rd}$ outlet patch correspond to $Q_0$ and $Q_{23}$, respectively, while those measured at other outlet patches correspond to $Q^s$ of the secondary branch ending at that patch.
The discrepancies w.r.t.\ $Q^{nom}$ are large only in correspondence of the lowest branches, but remain below $20\%$; they are indeed below $10\%$ for most $g$, thus ensuring that the flow velocity is of the appropriate order of magnitude to describe the effects of subsequent splitting.
Note that these mismatches are not an indication of the simulation accuracy, but only describe the preliminary calibration of the system.

\begin{figure*}
	\centering
	{\includegraphics[width=.25\textwidth]{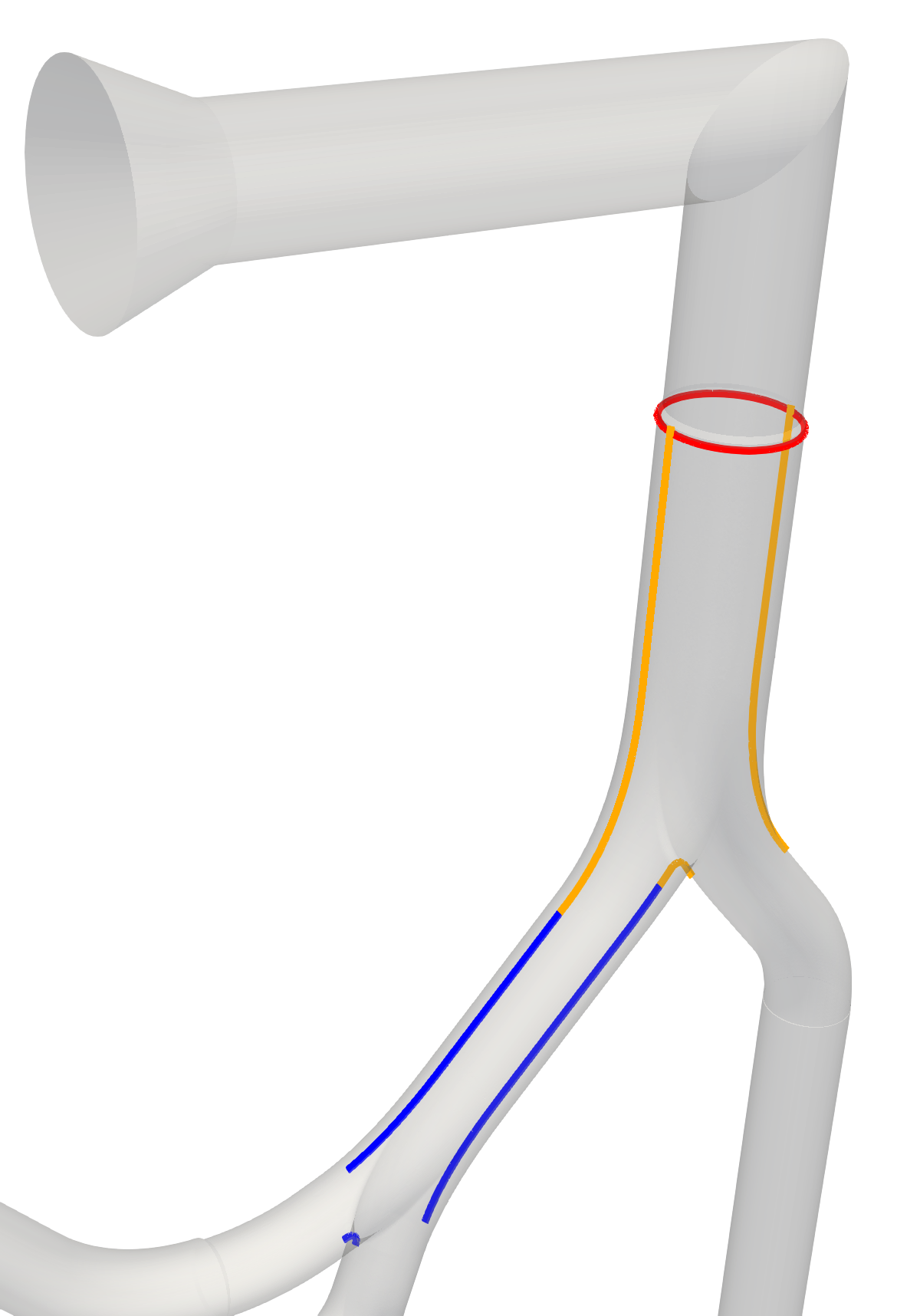}}
	{\includegraphics[width=.99\textwidth]{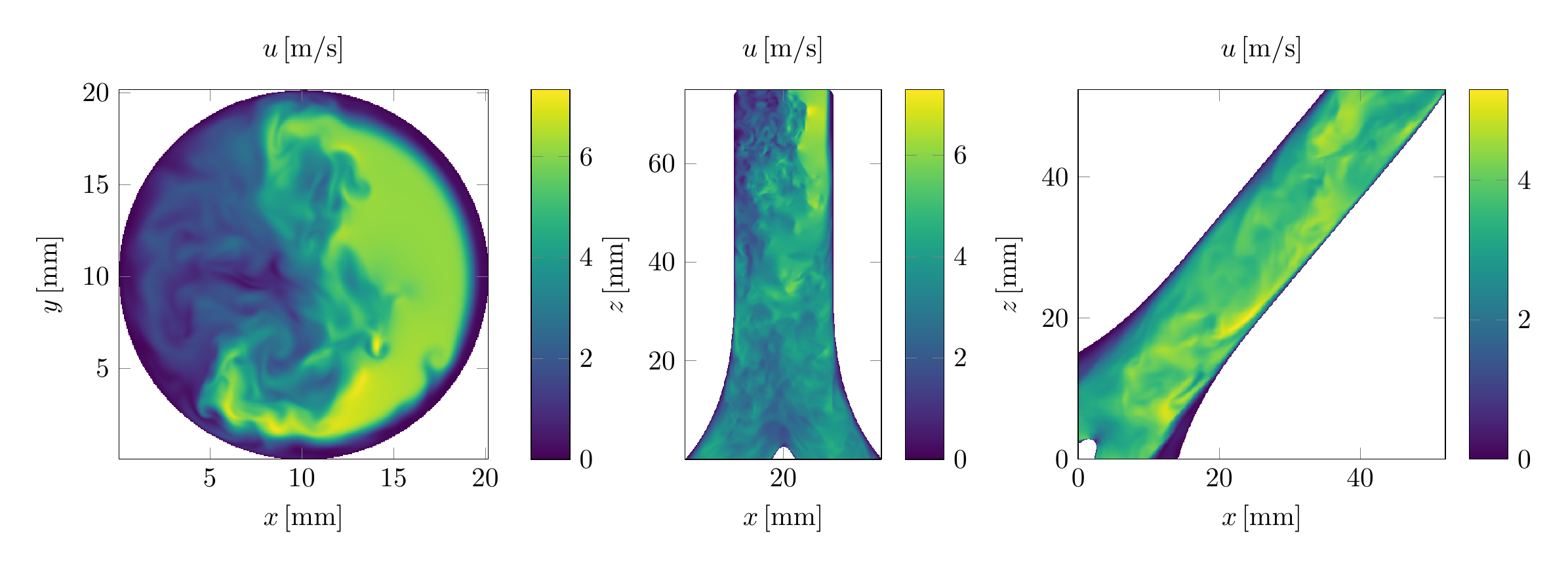}}
	\caption{Instantaneous velocity magnitude maps on the selected planes highlighted in red, yellow, and blue in the leftmost sketch. The first two sections are in the "trachea" of the model ($g=0$), the third one is in the first generation ($g=1$).}
\label{fig:inlet_condition}
\end{figure*}

With this choice of geometry and boundary conditions, no synthetic noise is introduced at the inlet, and a laminar-to-turbulent transition is not artificially triggered.
Flow separation naturally occurs at the sharp turn in the "throat" region of the model, and a chaotic regime emerges naturally at $g=0$, as illustrated in Fig.~\ref{fig:inlet_condition}.
In this figure, the instantaneous velocity magnitude is shown for a horizontal and a vertical section of the "trachea", i.e.\ $g=0$, and a vertical section of $g=1$. 
In the upper portion of $g=0$, two opposite regions of high and low air velocity following the $90^\circ$ turn of the idealized throat can be appreciated; the flow becomes more uniform before reaching the first bifurcation. The section at $g=1$ illustrates the chaotic flow that is entering the main branch of the model.

\subsubsection{Spatial and temporal resolution}
In DNS, the flow dynamics is solved without resorting to turbulence models.
However, the combination of different flow regimes, the fact that both instantaneous and mean flow are three dimensional, and the complex curvature effects that are inevitably present at bifurcations, make it impossible to define resolution criteria {\em a priori}. 
Owing to the geometrical complexity, a uniform distribution of the grid size in the Cartesian directions has been adopted to take maximum advantage of the IBM method. 
The progressive decrease of the local Reynolds numbers along the main branch, and the larger diameter of the secondary branches compared to the main branch from which they depart are two factors that help mitigate resolution requirements.
The number of grid points was selected to assure that at least $8$ points are present in the smallest channel, where a laminar solution is expected, requiring a grid spacing of $\Delta x = 52.5 \, \mu \rm m$. 
This resolution corresponds to $381$ grid points placed on the diameter of the largest section at $g=0$, which is comparable to that required for DNS of canonical pipe flow in the first generations \cite{pirozzoli-orlandi-2021} and finer in all following ones.
The total number of grid points is above one billion, at $\approx 1.4 \cdot 10^9$. 
Regarding time resolution, the simulation was carried out with CFL number lower than 1.5, which lead to a $\Delta t$, corresponding to the total time interval of the three-stage Runge--Kutta integration, of approximately $5.0 \cdot10^{-6}\,\rm s$. 

The significant computational cost prevents us from carrying out a conventional grid-dependency analysis, which would require one simulation with $\approx10^{10}$ grid points to be conclusive. 
However the analysis presented in Appendix~\ref{app:resolution} confirms that the space and time resolution of the present simulations are adequate.

\subsubsection{Statistical significance}
The simulation presented in this paper is started from the final flow field of the preliminary calibration runs. After discarding an initial period of $0.5\,\rm s$, flow statistics are accumulated for $1.5\,\rm s$ (i.e. for $\approx 3 \cdot 10^5$ time steps) in the entire domain, together with time series of flow quantities at preselected positions. 

The time interval for the accumulation of statistics was decided after examining the time-averaged flow rates for each generation computed for different sampling periods. 
In particular, the statistical uncertainty $\epsilon$ for the time-averaged flow rate $Q$, reported in Table~\ref{tab:summary}, was estimated following Russo \& Luchini~\cite{russo-luchini-2017}, confirming statistical convergence. 
Alongside $\epsilon$, the table reports another important quantity, namely the continuity error at each bifurcation, defined as the difference between the flow rate measured in the main branch before a bifurcation, and those in secondary branches, i.e.\ $\Phi_g=Q_{g}-(Q_{g+1}+Q^s_{g+1})$.
Note that $Q_0$ is the total averaged flow rate passing through the entire model, and the target ratio $Q_g/Q_{g-1}$ of the calibration phase is $0.5$ for all generations, except the last one; the last generation, $g=23$, does not have a parallel secondary branch by construction, and ideally has the same flow rate of $g=22$. 

\begin{table}[htp]
\centering
\begin{tabular}{ccccc}
$g$& $Q \,[\rm m^3/s]$ & $Q_{g}/Q_{g-1}$ & $\epsilon/ Q$ & $\Phi/Q$   \\ \midrule
$0$ &$9.97\cdot10^{-4}$ &-- &$7\cdot10^{-4}$ &$3\cdot10^{-5}$ \\ 
$1$ &$4.95\cdot10^{-4}$ &$0.497$ &$3\cdot10^{-3}$ &$7\cdot10^{-5}$ \\ 
$2$ &$2.51\cdot10^{-4}$ &$0.507$ &$3\cdot10^{-3}$ &$5\cdot10^{-5}$ \\ 
$3$ &$1.25\cdot10^{-4}$ &$0.497$ &$2\cdot10^{-3}$ &$4\cdot10^{-5}$ \\ 
$4$ &$6.25\cdot10^{-5}$ &$0.501$ &$1\cdot10^{-3}$ &$5\cdot10^{-5}$ \\ 
$5$ &$3.15\cdot10^{-5}$ &$0.504$ &$7\cdot10^{-3}$ &$4\cdot10^{-5}$ \\ 
$6$ &$1.57\cdot10^{-5}$ &$0.499$ &$5\cdot10^{-3}$ &$4\cdot10^{-5}$ \\ 
$7$ &$7.91\cdot10^{-6}$ &$0.502$ &$10\cdot10^{-3}$ &$5\cdot10^{-5}$ \\ 
$8$ &$3.91\cdot10^{-6}$ &$0.495$ &$9\cdot10^{-3}$ &$9\cdot10^{-5}$ \\ 
$9$ &$2.06\cdot10^{-6}$ &$0.526$ &$2\cdot10^{-2}$ &$7\cdot10^{-6}$ \\ 
$10$ &$9.99\cdot10^{-7}$ &$0.480$ &$3\cdot10^{-2}$ &$3\cdot10^{-4}$ \\ 
$11$ &$5.35\cdot10^{-7}$ &$0.530$ &$3\cdot10^{-2}$ &$1\cdot10^{-5}$ \\ 
$12$ &$2.57\cdot10^{-7}$ &$0.480$ &$9\cdot10^{-3}$ &$3\cdot10^{-4}$ \\ 
$13$ &$1.26\cdot10^{-7}$ &$0.491$ &$8\cdot10^{-3}$ &$1\cdot10^{-4}$ \\ 
$14$ &$6.29\cdot10^{-8}$ &$0.499$ &$1\cdot10^{-2}$ &$3\cdot10^{-4}$ \\ 
$15$ &$3.13\cdot10^{-8}$ &$0.497$ &$5\cdot10^{-3}$ &$9\cdot10^{-5}$ \\ 
$16$ &$1.60\cdot10^{-8}$ &$0.512$ &$4\cdot10^{-3}$ &$4\cdot10^{-5}$ \\ 
$17$ &$7.97\cdot10^{-9}$ &$0.498$ &$4\cdot10^{-3}$ &$7\cdot10^{-5}$ \\ 
$18$ &$3.96\cdot10^{-9}$ &$0.497$ &$4\cdot10^{-3}$ &$8\cdot10^{-6}$ \\ 
$19$ &$2.05\cdot10^{-9}$ &$0.518$ &$6\cdot10^{-3}$ &$1\cdot10^{-4}$ \\ 
$20$ &$9.38\cdot10^{-10}$ &$0.457$ &$3\cdot10^{-3}$ &$2\cdot10^{-4}$ \\ 
$21$ &$4.69\cdot10^{-10}$ &$0.500$ &$6\cdot10^{-3}$ &$2\cdot10^{-5}$ \\ 
$22$ &$2.46\cdot10^{-10}$ &$0.524$ &$2\cdot10^{-2}$ &$9\cdot10^{-4}$ \\ 
$23$ &$1.93\cdot10^{-10}$ &$0.784$ &$8\cdot10^{-3}$ &$4\cdot10^{-4}$ \\ 
 
\end{tabular}
\caption{Evolution of mean flow rate $Q$, flow splitting $Q_g/Q_{g-1}$, relative uncertainty $\epsilon/Q$ of the mean flow rate, and relative continuity error $\Phi/Q$ along the generations.}
\label{tab:summary}
\end{table}
Throughout all branches, the relative statistical uncertainty of the flow rate is at most of $\approx 3\%$. 
The relative continuity error is always lower than 0.01\% for the first 9 generations; its largest value is a relative $0.1\%$ at generation $22$, where  $\Phi \approx 10^{-14} \, \rm m^3/s$.

\section{Results and discussion}
\label{sec:results}

We start examining the different flow regimes in various regions of the model; then, we address how transitioning between regimes affects the time dependence of relevant physical variables; we build on these findings to link the characteristic time scales in the first generation to those in deeper generations; we discuss how flow rate fluctuations propagate across the main branch of the model; lastly, we address the limitations of this work caused by a departure from physiological conditions.

\subsection{Flow regimes across generations}
\label{sec:res_regimes}
\begin{figure}[htp]
\centering
\includegraphics[width=0.75\textwidth]{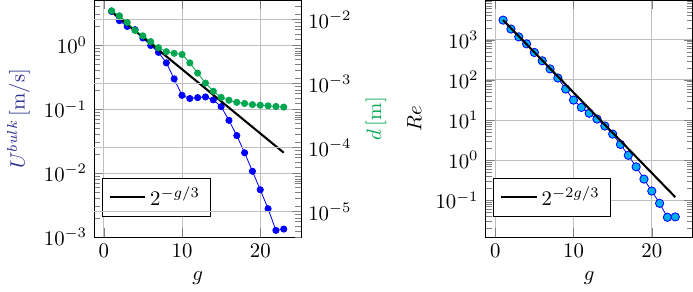} 
\caption{Evolution across generations of the local diameter $d$, the local bulk velocity $U^{bulk}$ (left), and the local Reynolds number (right).}
\label{fig:ReI}
\end{figure}

We begin by introducing the local Reynolds number $Re = U^{bulk} d / \nu = 4 Q / (\pi d \nu)$, where $d$ is the local diameter and $U^{bulk}$ the corresponding bulk velocity, i.e.\ $Q = \pi U^{bulk} d^2/4$.
The evolution of $d$, $U^{bulk}$ and $Re$ across generations is shown in Fig.~\ref{fig:ReI}. 
With the selected flow partitioning, $Q$ decreases by a factor approximately 1/2 when passing from one generation to the next.
The model geometry is such that, down to $g=7$, diameter and velocity decrease at the same rate, which is $2^{-1/3}$. 
For $g>7$, the diameter still monotonically decreases along with the main branch, but with a variable rate of change. 
In particular, $d_g/d_{g-1}>2^{1/2}$ for $11<g<13$, causing a plateau for $U^{bulk}$ in this intermediate range, whereas $U^{bulk}$ returns to a fast drop rate for $g>13$.  
The combination of $d$ and $U^{bulk}$ is such that the local Reynolds number monotonically decreases through the main branch, from $Re_0 \approx 3000$ at the beginning to $Re_{23}\approx 0.03$ in the last generation. For most of the system, $Re$ is seen to follow closely the law $2^{-2g/3}$ that derives from the combined decrease of $d$ and $U^{bulk}$.

\begin{figure*}[t]
\centering
\includegraphics[width=\textwidth]{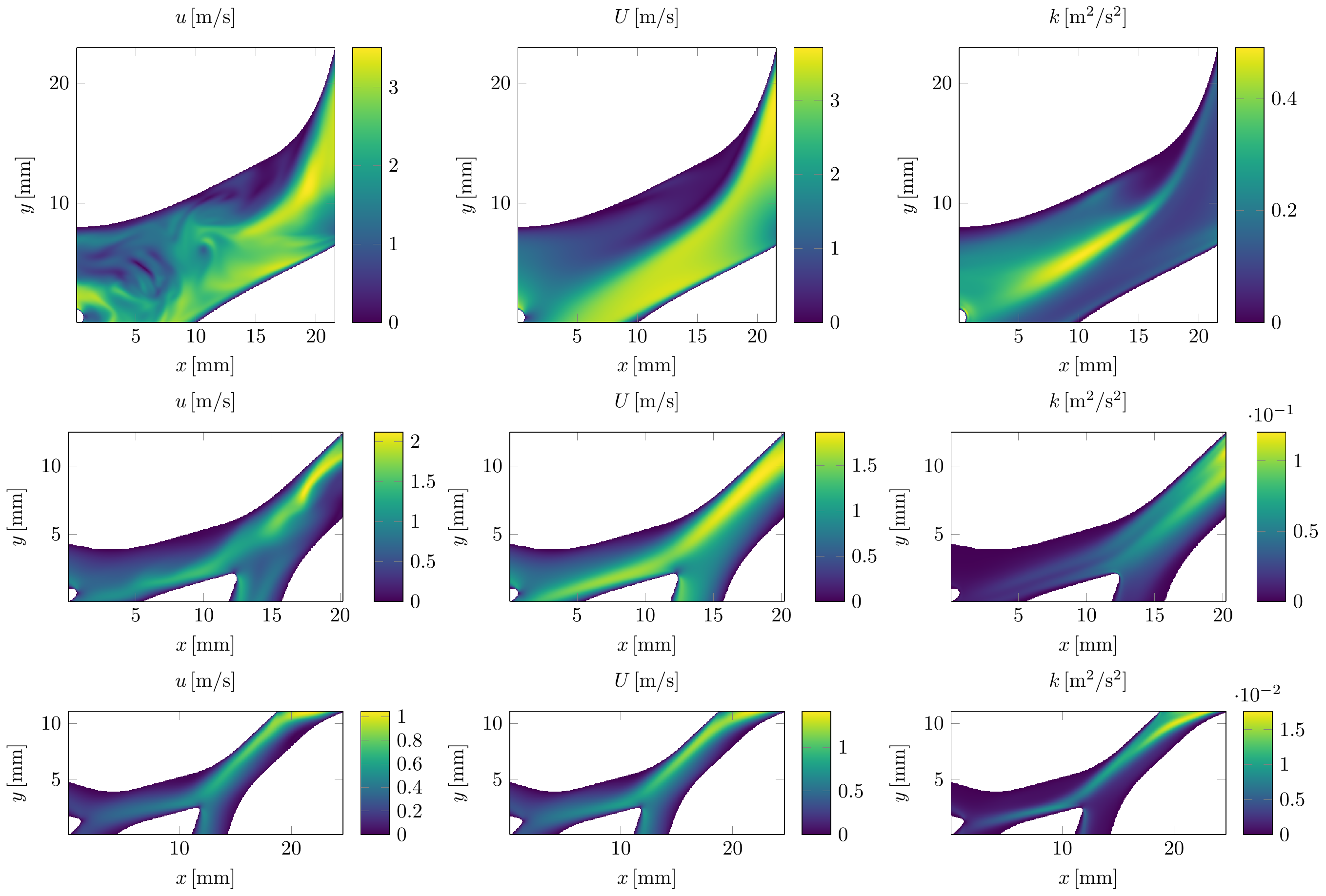}
\caption{Magnitude of the instantaneous $u$ (left) and mean $U$ (centre) velocity, and fluctuation kinetic energy $k$ (right) for $g=4$ (top), $g=7,8$ (middle) and $g=9,10$ (bottom). The Cartesian coordinates are shifted so that the bottom left corner is $(0,0)$.}
\label{fig:generations}
\end{figure*}

We shall now examine a set of flow visualizations, to discuss the qualitative changes of the flow while $Re$ evolves along the main branch.
In Fig.~\ref{fig:generations}, the instantaneous and time-averaged velocity magnitude, denoted with $u$ and $U$, respectively, are shown in three planar sections of the domain, at $g=3$ (representative of the portion of main branch with Reynolds numbers high enough to sustain turbulent fluctuations) and $6 \le g \le 10$. The figure also plots the quantity $k = {\rm Mean}[\bm{u}'\cdot\bm{u}']/2$, that will be referred to in the following as \textit{fluctuation kinetic energy}, a general expression that is appropriate for both turbulent and unsteady laminar flows.

At $g=3$, the flow qualitatively resembles a typical turbulent flow, with high values of $k$, and a mean field that differs from the instantaneous field. In the latter, coherent structures advected through the duct can be recognized. 
Wall-curvature effects are also evident, as a region of low flow velocity appears in the upper portion of the duct, and a corresponding region of high flow velocity is present in the lower portion. 
Since the Reynolds number is large enough, the shear layer that delimits the two regions, combined with the disturbances convected from upstream, is unstable, oscillates and breaks into smaller structures.
Similar patterns have been observed in more realistic geometries, with non-planar bifurcations~\cite{bernate-etal-2017}.

As the Reynolds number decreases, for $g=7$ and $8$, the smaller-scale fluctuations disappear but the instantaneous field remains visibly different from the mean field, confirming the residual presence of turbulence.
Eventually, however, for $g=9$ and $10$, it becomes difficult to distinguish any structure convecting through the main branch, and the distinction between $u$ and $U$ becomes less evident, although asymmetry and secondary flows caused by wall curvature are still present. 
In fact, instantaneous and mean velocity fields still differ, but the time dependency is a simple modulation rather than a consequence of convecting flow structures. 
The transition from a chaotic or turbulent-like regime to an unsteady laminar one is effectively represented by the fluctuation kinetic energy field. 
At $g=3$, $k$ is very different from $U$, and peaks in the regions with large velocity gradients. 
Moving towards higher $g$, the peaks of $k$ progressively tend to coincide with those of $U$, confirming that time dependency here manifests as scaling of the velocity. 
From a qualitative perspective, this transition is in agreement with observations in more complex models that include intermediate generations~\cite{jing-etal-2023}. 
Discrepancies are limited to the generation where a change of regime takes place or to details of the solution, which are to be expected, given the differences in geometries and flow rates.  

\begin{figure*}[t]
\centering
\includegraphics[width=\textwidth]{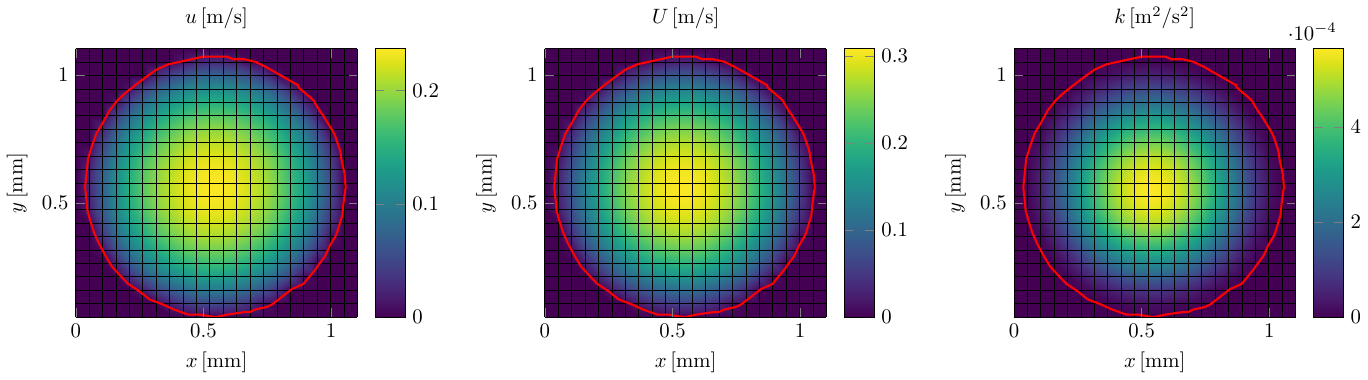}
\caption{From left to right: instantaneous velocity magnitude $u$, mean velocity magnitude $U$, and fluctuation kinetic energy $k$ at $g=13$, sampled at the same time. Cartesian coordinates are shifted so that the bottom left corner is $(0,0)$. The black grid and the red line indicate the position of grid points and the actual boundary.}
\label{fig:gens2}
\end{figure*}

We finally consider a cross-section of generation $g=13$, shown in Fig.~\ref{fig:gens2}.
The Cartesian grid at this generation may appear relatively coarse w.r.t.\ the duct size, but the presence of the wall is properly taken into account through the IBM correction.
Here, the Reynolds number is low enough to fully suppress secondary flows. 
In fact, both $u$ and $U$ are virtually indistinguishable from the laminar parabolic solution in a straight cylindrical pipe.
Nonetheless, the fluctuation kinetic energy presents relatively high values. 
In particular, its intensity $\sqrt{2k/3}/U$, which is a parameter often used in CFD, is $\approx 4\%$ at the pipe centre, a value that is fully compatible with a turbulent regime. 
However, the spatial distribution of $k$ does not resemble at all that expected in a turbulent pipe or other classical wall-bounded flows, where a pronounced near-wall maximum is present; here, to the contrary, the profile of $k$ is parabolic, akin to that of the velocity.
The same picture is observed for all generations below $g=13$. 

We will now adopt a more precise nomenclature for flow regimes in different generations. 
Hereafter, the term ``turbulent" will refer to generations where time dependence is not a simple rescaling, for $g \lesssim 8$; the term ``laminar'', instead, is used for $g \gtrsim 9$. 
For laminar flow, a further distinction can be introduced, considering whether secondary flows are present or not. 
For generations $g \gtrsim 13$, where streamlines mostly remain parallel to the walls and secondary flows are negligible, we will use the term ``Stokes flow".

\subsection{Time scales}
\label{sec:res_history}
We now focus on the time-dependent fluctuations of flow rate, velocity magnitude, and pressure around their time-averaged values, i.e. $q'$, $u'$, and $p'$.
While flow rates are integrated over a cross-section, velocity and pressure are pointwise-varying quantities, which we sample approximately at the central point of the geometrical centre line at each generation. 

\begin{figure*}[t]
\includegraphics[width=\textwidth]{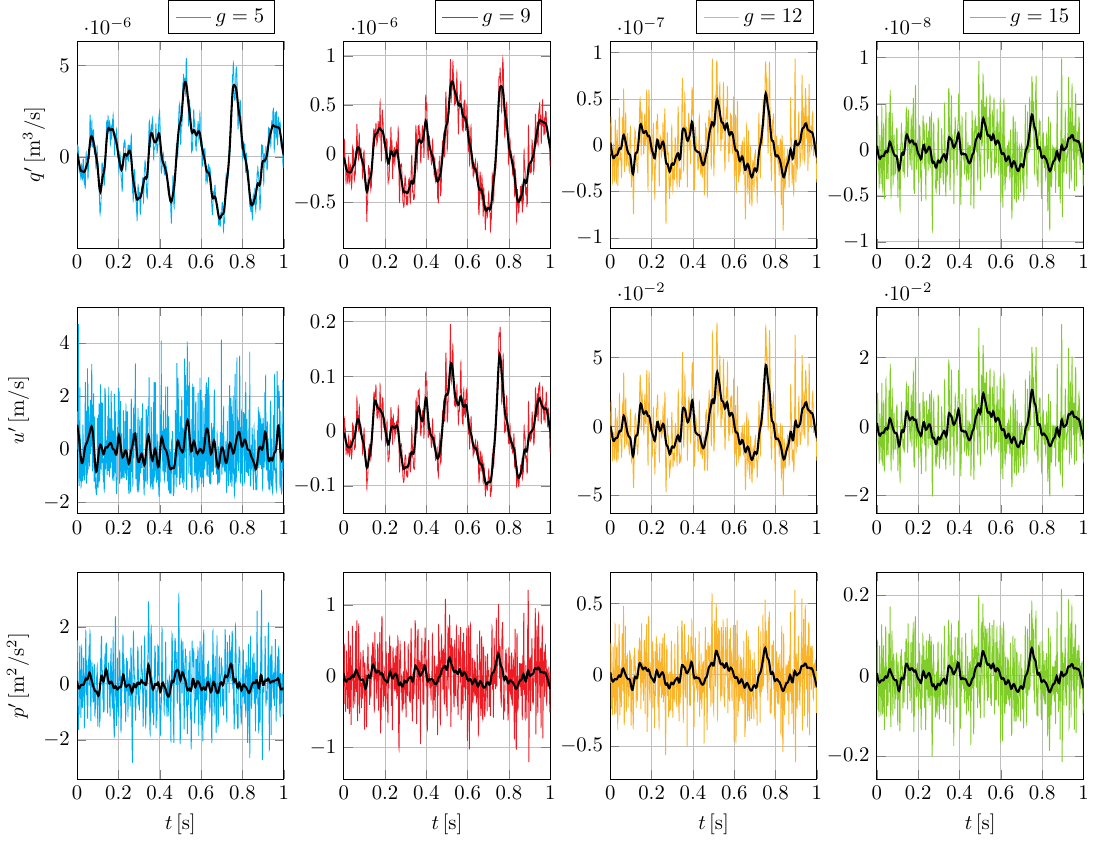}
\caption{Time histories for the flow rate (top), fluctuations of centre line velocity (middle) and fluctuations of centre line pressure (bottom), for generations $g=5,9,12,15$. Each panel includes a low-pass filtered version of the signal (black line).}
\label{fig:probes}
\end{figure*}

The time histories for $g=5$, $9$, $12$, and $15$ are shown in Fig.~\ref{fig:probes}. 
Generation $5$ is taken as a proxy of all generations above, which exhibit a similar time history, and generation $15$ is equivalent to those below.
To highlight the distinction between fast and slow time scales, the fluctuations of the three variables are presented alongside with their filtered version where a low-pass filter is used to remove fast scales. 
Filtered fluctuations are denoted with a tilde, as in $\tilde{q}$ for the flow rate. 
A Gaussian filter with cut-off frequency $f_0 = 5 \, {\rm Hz}$ is used, where $f_0$ is comparable to the large and slow fluctuations at $g=1$. 
We note that the qualitative results presented here are unchanged after different choices of $f_0$ at least up to $\approx 10 \, {\rm Hz}$; lower values of $f_0$ are not appropriate, since the total time history is only 1.5 seconds.
At $g=5$ and $9$, the major flow rate fluctuations are not suppressed by the low-pass filter, as they are slower than the time scale $1/f_0$, similar to those entering the main branch. 
This is in agreement with the qualitative observations in the previous section. 
Indeed, as long as advection remains relevant, coherence is retained through the main branch, and large-scale coherent structures in the turbulent flow leave their footprint on the temporal evolution of the flow rate. 
Moving towards lower generations, this modulation becomes less evident, and faster flow rate oscillations emerge.
More precisely, the filtered signal $\tilde q$ that retains the slow time scales is fairly similar in shape through all generations, but slow time scales do not associate with fluctuations of large amplitude.
Contrary to $q'$, the signal of $u'$ is more evidently affected by the turbulent-to-laminar transition that occurs at intermediate generations. 
Both $u'$ and $\tilde u$ are poorly correlated with $q'$ at $g=5$, which is expected for chaotic flows, while they follow much more closely $q'$, at all generations downstream, irrespective of the progressive attenuation of $\tilde q$.
This is the expected behaviour for the velocity in flow regimes where time dependence manifests itself as rescaling of the solution. 
Lastly, since pressure in incompressible flows obeys the Poisson equation, with velocity on the right-hand-side, pressure structures are much larger than velocity structures, and the pointwise time history of the two is apparently uncorrelated from one another, as long as advection remains relevant. 
However, as advection becomes progressively less relevant w.r.t.\ viscous diffusion in deeper generations, history effects become less important to determine the local state of the flow, and $p'$ becomes more tightly linked with $q'$ and, by extension, $u'$.
Eventually, in the Stokes regime, flow rate or pressure (differences) can uniquely determine the local solution.

\begin{figure}
\centering
\includegraphics[width=0.5\textwidth]{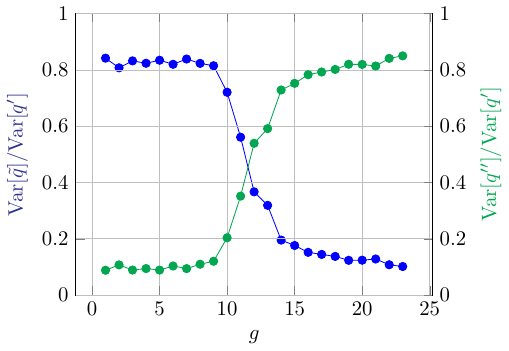} 
\caption{Relative variance of the slow (left axis) and fast (right axis) fluctuations of flow rate for a cut-off frequency of $f_0=5{\rm Hz}$.}
\label{fig:low_pass}
\end{figure}

We can now focus on the attenuation of flow rate fluctuations with different time scales.  
To distinguish between slow and fast time scales, let us introduce a triple decomposition based on the low-pass filtered flow rate: 
\begin{equation}
\begin{split}
& q''(t)= q'(t)-\tilde q(t)\\
& q(t) = Q + \tilde q (t) + q''(t) \,.
\end{split}
\label{eqn:triple_dec_q}
\end{equation}
This triple decomposition separates the time-dependent flow rate $q$ into the time-mean value $Q$, the low-pass filtered signal $\tilde q(t)$, and the difference $q''(t)$ between $q'(t)$ and $\tilde q(t)$.
To provide a quantitative measure of the fluctuations intensity associated to different time scales, we consider their variance normalized with the variance of the entire signal, i.e.: ${\rm Var}[\tilde q] / {\rm Var}[q']$ and ${\rm Var}[q''] / {\rm Var}[q']$. 
These quantities are shown in Fig.~\ref{fig:low_pass} for all generations, and exhibit a regular evolution along the main branch. 
In particular, the normalized variance of the slow fluctuations remains approximately constant and around $0.8$ down to $g=9$, in the regions characterized by a turbulent regime.
In this portion of the main branch, therefore, not only slow fluctuations have proportionally higher amplitude than fast fluctuations, but also slow and fast time scales of the flow rate are similarly affected moving from one generation to the next.
Below $g=9$, however, the variance of the slow fluctuations rapidly decreases, and drops below 0.2 at $g>14$ .
The variance of the fast fluctuations exhibits the opposite and complementary trend, becoming larger than the slow one from $g=9$ downwards.

\begin{figure}
\centering
\includegraphics[width=0.75\textwidth]{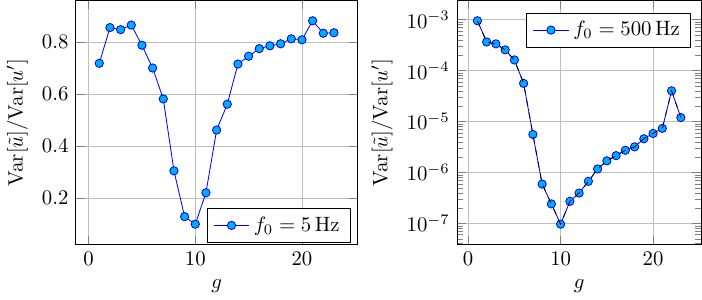} 
\caption{Evolution of the relative variance of slow and fast velocity fluctuations, obtained from low-pass filtering the signal of $u'$ with $f_0=5\,\rm Hz$ (left) and $f_0 = 500\,\rm Hz$ (right).}
\label{fig:filtered}
\end{figure}

From the picture drawn so far and the visualisation in Fig.~\ref{fig:probes}, one might get the false impression that the progressively lower Reynolds number does not affect fast fluctuations. 
This is not the case. To clarify this point, we now apply the triple decomposition to velocity fluctuations, and focus on the relative variance of the slow velocity fluctuations, namely ${\rm Var}[\tilde u] / {\rm Var}[u']$. 
To highlight the different evolution of fast and slow time scales, we consider two cut-off frequencies, i.e. $f_0=5 \,\rm Hz$ and $f_0 = 500 \,\rm Hz$, as illustrated in Fig.\ref{fig:filtered}.
Note that the higher cut-off frequency $f_0=500\,\rm Hz$ is chosen here simply to be much larger than $f_0$, so that the associated relative intensity at the highest generation is $\approx10^{-3}$.
With $f_0=5\,\rm Hz$, all time scales are "fast" and filtered out but those associated with large flow rate oscillations in the upper generations. 
Consistently with the time history, ${\rm Var}[\tilde u] / {\rm Var}[u']$ is between $0.7$ and $0.85$ both at generations with a turbulent or non-turbulent regime, i.e.\ before $g=9$, and at generations in the Stokes regime with fast fluctuations, i.e.\ after $g=15$. 
The relative oscillation intensity decreases only at intermediate $g$, reaching its minimum value of $0.1$ at $g=13$, where two conditions are simultaneously met: $u'$ is following $q'$, but $q'$ still exhibits slow fluctuations with high amplitude. 
Contrary to what observed for $f_0=5\,\rm Hz$, when the faster cut off frequency $f_0=500 \,\rm Hz$ is considered, the relative fluctuation intensity immediately starts to decrease, as the range of active scales becomes narrower with decreasing $Re$. 
In particular, ${\rm Var}[\tilde u] / {\rm Var}[u']$ at $g=6$ is already one order of magnitude lower than that at $g=1$, and eventually reaches a minimum of $\approx 10^{-7}$ at $g=10$. 
After that point, including generations where the Stokes regime is established, the relative variance increases again, albeit remaining much lower than in the turbulent regime. 
In particular, while ${\rm Var}[\tilde u] / {\rm Var}[u']$ for $f_0=5\,\rm Hz$ has already reached $0.7$ at $g=14$, ${\rm Var}[\tilde u] / {\rm Var}[u']$ for $f_0=500\,\rm Hz$ at $g=23$ is still two order of magnitudes lower than in $g=1$.

The increase of ${\rm Var}[\tilde u] / {\rm Var}[u']$ in the portion of main branch where $Re<1$ cannot be associated, as it would for a turbulent flow, to a larger range of active scales. 
On the contrary, the local quantity $u'$ follows the same evolution of $q'$, and a higher fluctuation intensity for the velocity is directly connected to that of the flow rate. 
We can thus conclude that the progressive reduction of Reynolds number in the main branch does indeed lead to a suppression of the fast scales down to the intermediate generations. 
Further down, however, where velocity and flow rate become locked in, fast scales can become more important than the value of the local Reynolds number would imply.

The results presented thus far highlight that, even though the progressively lower $Re$ is associated to a transition towards the Stokes regime, where coherent structures in the sense commonly intended for turbulent flows are absent, flow rate fluctuations still exhibit relatively fast time scales in all generations.  
In the next section, we discuss whether it is possible in these circumstances to identify a link between the regime at $g=1$ and that in following generations. 

\subsection{Connection between upper and lower generations}
\label{sec:res_scales}
We now consider the local Womersley number, $Wo$, which is typically employed to measure the relative importance of inertia and viscous diffusion in the context of pulsatile flows, where extracting a characteristic frequency is immediate. 
In our case, as the breathing is steady, there is no pulsatile flow and the Womersley number as usually defined is not a meaningful parameter. 
It is, however, still possible to introduce a definition based on a characteristic timescale, $\tau$, i.e.:\ $Wo = d / \sqrt{\nu \tau}$. 
In this way, the Womersley number serves as a dimensionless frequency that measures the unsteadiness of physical quantities. 
Hereafter, we adopt the time delay of the first zero-crossing of the temporal autocorrelation function as definition of $\tau$ (for a perfectly sinusoidal time signal, $\tau$ would be exactly $1/4$ of the period).
The most relevant physical quantity to characterize the flow at a given generation is the local flow rate, whose corresponding timescale and Womersley number are denoted with $\tau^q$ and $Wo^q$, respectively.
We intend to ascertain whether, notwithstanding the relatively complex evolution of the flow along the main branch, an estimate for the local $Wo^q$ at each generation can be arrived at solely based on flow properties at $g=1$. We denote the estimate of local Womersley as $Wo^*$. 

\begin{figure*}
\centering
\includegraphics[width=0.99\textwidth]{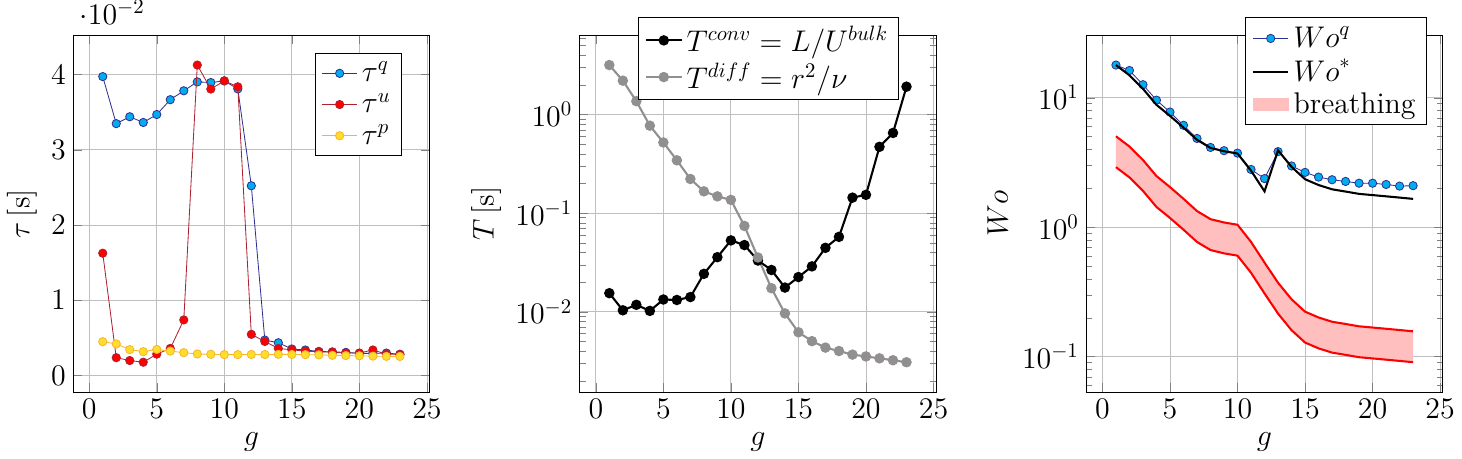} 
\caption{Left: timescale $\tau$ for the quantities $q'$, $u'$ and $p'$. Center: convective and diffusive timescales. Right: Womersley number, comparison between the observed $Wo$ (blue dots), estimated $Wo$ (black line), and $Wo$ computed for breathing cycles with period between $2$ and $6$ seconds (red area).}
\label{fig:tau}
\end{figure*}

Before a definition of this estimate can be formulated, it is necessary to identify connections between time scales in different generations. 
We begin with examining the timescales $\tau^q$, $\tau^u$ and $\tau^p$, shown for each generation in Fig.~\ref{fig:tau}. 
These quantities are introduced as tools to highlight the transitions between regimes with long time scales associated with large or small fluctuations. 
Their general trend is in agreement with the qualitative observations from the time history: $\tau^q$ remains in the interval $[3,4]\cdot10^{-2}\,\rm s$ at $g<12$, and drops by approximately one order of magnitude at $g>12$; $\tau^u$ is lower than $\tau^q$ until $g=8$, and then follows $\tau^q$ for the generations below; finally, $\tau^p$ remains approximately one order of magnitude lower than $\tau^q$ up to $g=13$. 
Note that these values corroborate the choice made in the previous section of $f_0=5\,\rm{Hz}$ and $f_0=500\,\rm{Hz}$ as representatives of ``slow" and ``fast'' frequencies, respectively.

We note that the flow is turbulent in generations with $g<8$. 
The local $Re$ is not high enough to sustain fully-developed turbulence as traditionally intended, but there is still a certain degree of scale separation in the flow, that is here apparent in the difference between $\tau^q$ and $\tau^u$. 
The laminar regime holds for $8 \leq g < 11$. Here, scale separation is reduced to the minimum, resulting in $\tau^u \approx \tau^q$, but some coherence transmitted from higher generations still exists, so $\tau^q_{8-11} \approx \tau^q_{1-7}$. 
Lastly, the Stokes regime arises for $g \geq 12$. Here, spatial coherence is lost and the modulation by large time scales disappears, so $\tau^q$, $\tau^u$, and $\tau^p$ are short and similar to one another.

Even though describing the complex transition between regimes in the main branch through $\tau$ is inevitably a simplification, the results shown in Fig.~\ref{fig:tau} (left) suggest a path to link $g=1$ to all the generations below.
In fact, before the transition to the Stokes regime, $\tau^q_1$ provides a good estimate for all $\tau^q_g$; similarly, since $\tau^p$ exhibits relatively small variations in the entire bronchial tree, $\tau^p_1$ provide a reasonable guess for $\tau^q_g$ after the Stokes regime is established. 

We can now take advantage of our knowledge of the model geometry and average splitting coefficients, which are considered as known data, to estimate where the transition to the Stokes regime takes place.
Let us define two time scales that can help identifying the generation where spatial coherence is lost.
The first one is a local convective time scale: $ T^{conv} = L /  U^{bulk}$, where $L$ is the length of the centre line segment corresponding to generation $g$, and $U^{bulk}$ is the mean bulk velocity. 
The second one is a diffusion time scale built with (half) the diameter $d$ and the fluid viscosity, i.e. $T^{diff} = d^2/4 \nu$. 
The convective time scale quantifies the travel time across generation $g$, whereas the diffusion time scale estimates the time necessary for diffusion to smear localized disturbances over the cross-section.
Note that $T^{conv}$ depends only on $Q_0$, $Q_0/Q_g$, and the geometry of the model, while $T^{diff}$ is determined by geometry alone. 
The evolution across the generations of these two time scales is shown in Fig.~\ref{fig:tau} (center). 
The convective time scale has a relatively complex behaviour, while the diffusion time scale rapidly decreases in the first portion of main branch, following the evolution of $d$. 
It is found that the Stokes regime is established when $T^{diff} \approx T^{conv}$, a condition that takes place at $g=12$. 

We are finally in the position of defining the estimate $Wo^*$.
Two cases exist: i) if $T^{diff} < T^{conv}$, before the Stokes regime is established, the relevant time scale is the flow rate time scale at $g=1$, i.e.\ $\tau^{q} \approx \tau^q_1$ and $Wo^*$ is built with $\tau_1^q$; ii) if $T^{diff} > T^{conv}$, after the Stokes regimes is established, the relevant time scale is the pressure time scale at $g=1$, i.e.\ $\tau^{q} \approx \tau^p_1$, and $Wo^*$ is built with $\tau_1^p$. 
A comparison between $Wo^*$ and the measured local $Wo^q$ is shown in Fig.~\ref{fig:tau} (right), together with the Womersley number computed with $\tau$ from the autocorrelation functions of sinusoidal signals with periods $2$, $4$, and $6\,\rm s$, which corresponds approximately to fast, average, and slow breathing \cite{finlay-2019}. 
The comparison between these quantities should be regarded as qualitative only, since fluctuations with characteristic frequency $Wo^q$ (or $Wo^*$) do not cause a change of sign of the flow rate, contrary to those caused by the breathing cycle; nevertheless, it helps clarifying the separation between breathing and fluctuations time scales.
One notices that the local $Wo^q$ and its counterpart based on breathing frequencies are quite different, in terms of both values and trend: $Wo^q$ is always larger, and does not exhibit a rapid decrease in intermediate generations. 
In this region of the bronchial tree, in fact, the sudden drop of $\tau^q$ counterbalances the diameter reduction, illustrating that flow rate fluctuations in our test case lead to much faster frequencies in deep generations than those that can be associated to the breathing cycle.
On the other hand, the comparison of $Wo^q$ and $Wo^*$ confirms that $\tau^q_1$ gives a good estimate of the level of unsteadiness before the Stokes regimes is established, while $\tau^p_1$ is a more appropriate time scale beyond that point.
The agreement between $Wo^q$ and $Wo^*$ is not perfect in the last generations, where the local $\tau^q$ is still lower than $\tau_1^p$, but nonetheless quite remarkable considering the dramatic change of flow regime across the main branch. 

Besides the obvious limitations of such a rudimentary model to estimate $Wo$ across generations, the considerations above emphasize the major result that relatively fast and small fluctuations introduced at the top of the bronchial tree do remain important throughout the entire main branch. 
In fact, they may be even more important than large fluctuations.
While this is surprising at first, it is actually quite a reasonable outcome, given the large range of flow rates across generations. 
Oscillations in the flow rate by one percent at a given generation are in fact of the order of magnitude of the total flow rate 6-7 generations below. 

\subsection{Propagation of flow rate fluctuations}
\label{sec:res_flow_fluctuations}
Here we discuss how the amplitude of flow rate fluctuations propagates through the bronchial tree. 
So far, the ratio $Q_{g+1} / Q_g \approx 1/2$ of average flow rates at a bifurcation has been treated as a property of the study case. 
We shall now take a different viewpoint, and consider the local instantaneous flow rate split coefficient, $s(t)$, defined so that: $q_{g+1}(t) = s_g(t) q_g(t)$. 
Adopting the usual notation for mean and fluctuations, $s(t) = S + s'(t)$. 
We can now connect flow rates and split coefficients:
\begin{equation}	
	Q_{g+1} = {\rm Mean}[(S_g + s'_g)(Q_g+q'_g)] = S_g Q_g + {\rm Cov}[s_g,q_g]\,,
\end{equation}
where ${\rm Cov}[s,q]$ denotes the covariance.
From this expression, we can obtain that:
\begin{equation}
	\frac{Q_{g+1}}{Q_g} = S_g + \frac{{\rm Cov}[s_g, q_g]}{Q_g}\,,
\end{equation}
which shows that $Q_{g+1} / Q_g \approx S_g$ in the limit of small fluctuations or if $s_g$ and $q_g$ are not correlated.

We can now examine the connection between the statistical properties of $s(t)$ and $q(t)$.
In steady conditions, $q_g(t)=Q_g$ and $S_g=Q_{g+1} / Q_g$. In this case, $q_{g+1}$ is also constant, and we simply have:
\begin{equation}
\begin{cases}
	Q_{g+1} = S_g Q_g \\
	{\rm Var}[q_{g+1}] = 0
\end{cases}\,.
\end{equation}
For the unsteady case of interests here, the simplest possibility is when the split coefficient is constant: $q_{g+1}(t)$ equals to $q_{g}(t)$ multiplied by the constant $S_g$, and we have: 
\begin{equation}
\begin{cases}
	Q_{g+1} = S_g Q_g \\
	{\rm Var}[q_{g+1}] = S_g^2 {\rm Var}[q_{g}]
\end{cases}\,.
\label{eqn:variance_with_qt}
\end{equation}
This is the condition at which fluctuations of a given intensity w.r.t.\ the local mean flow rate are simply transmitted with the same intensity to the next generation.
Although the assumption of constant $s$ may seem fairly unrealistic, eq.~\eqref{eqn:variance_with_qt} serves as an important reminder of how semi-steady conditions are not compatible with an attenuation of flow rate fluctuations.
While velocity fluctuations can be dissipated if the Reynolds number is low enough, flow rate fluctuations do maintain their relative intensity in this case.

When both $q$ and $s$ are functions of time, we simplify the problem by assuming for both quantities small fluctuations w.r.t.\ their mean values. 
We can then write $q_{g+1}$ as a first-order expansion in a neighbourhood of its mean value:
\begin{equation}
	q_{g+1} \approx S_g Q_g + (s_g-S_g) Q_g + S_g (q_g-Q_g) \,.
\end{equation}
From this expression, the properties of the variance can be used to derive an approximation of ${\rm Var}[q_{g+1}]$:
\begin{equation}
\begin{split}
	{\rm Var}[q_{g+1}] &\approx{\rm Var}[S_g Q_g + (s_g-S_g)Q_g + S_g (q_g-Q_g)]\\
	& \approx {\rm Var}[ s_g Q_g + S_g q_g  ] \\
	& \approx Q_g^2 {\rm Var}[s_g] + S_g^2  {\rm Var}[q_g]  + 2 S_g Q_g {\rm Cov}[s_g, q_g] \, .
\end{split}
\end{equation}
Mean and variance of the flow rates after a bifurcation can then be written:
\begin{equation}
\begin{cases}
	Q_{g+1} \approx S_g Q_g \\
	{\rm Var}[q_{g+1}] \approx Q_g^2 {\rm Var}[s_{g}] + S_g^2 {\rm Var}[q_{g}] + 2 S_g Q_g {\rm Cov}[s_g, q_g]
\end{cases}\,.
\label{eqn:variance_with_qt_and_at}
\end{equation}
This is an interesting expression, because the last term ${\rm Cov}[s_g, q_g]$ can be of either sign, and thus contributes, in principle, to an attenuation of fluctuation intensity along the main branch.

Several observations are in order. 
First, we should quantify the errors introduced by the first-order expansion of stochastic variables. 
The prediction of fluctuations intensity obtained with eq.~\eqref{eqn:variance_with_qt_and_at} is compared against DNS in \S~\ref{app:validity_expansion}, showing excellent agreement and confirming the soundness of this approach.
Second, it is natural to conceive a connection between the flow regime at a given generation and ${\rm Var}[q]$ or ${\rm Var}[s]$. 
Turbulent flows in ducts exhibit time-dependent losses, which can lead to a time dependence for hydraulic resistance.
Flow splitting will then also be time-dependent in a bifurcation where the flow is turbulent in the two daughter branches.
However, for the covariance ${\rm Cov}[s, q]$ to be non-zero such unsteadiness is not enough, and a geometrical bias is required.
Moreover, ${\rm Cov}[s, q]$ has opposite sign on the two daughter branches after a bifurcation, since $q$ is the same for both, and the sum of $s$ for the two branches must be $1$. 
It is then more appropriate to say that equations~\eqref{eqn:variance_with_qt_and_at} allow a sort of redistribution of fluctuation intensity of the flow rate, rather than a true attenuation.
Lastly, we can point out that, even though there is a formal symmetry in \eqref{eqn:variance_with_qt_and_at} in how $s_g$ and $q_g$ contribute to ${\rm Var}[q_{g+1}]$, these variables possess a significantly different range of variations. 
Flow splitting coefficients are pure numbers, which can be larger than $1$ only in the presence of reverse flow in one branch, and should be fairly similar and with a low variance in all generations, in particular in the statistically steady conditions considered here. 
To the contrary, $q$ spans at least $6$ order of magnitudes, and $q'(t)$ exhibits significant differences across the various generations.

Some additional simplifications can be introduced in equations~\eqref{eqn:variance_with_qt_and_at}, to extract at least a qualitative indication of how splitting fluctuations affect flow rate fluctuations.
All averaged flow rates, $Q$, and the variance for the first generation, ${\rm Var}[q_1]$, are taken as known data; furthermore, all generations are given the same values ${\rm Var}[s_g] = A$ and ${\rm Cov}[s,q]$ is assumed to be zero. 
This simplified form of eq.~\eqref{eqn:variance_with_qt_and_at} cannot take into account possible attenuation or amplification caused by the covariance term. 
There is, however, another consequence of the assumptions we just introduced that may be less apparent. 
Taking for simplicity $Q_g = Q_0/2^g$ and $S_g = 0.5$, which are fairly close to the measured $Q_g$ and $S_g$ in our case, we obtain the following recursive formula:
\begin{equation}
{\rm Var}[q_{g+1}] = \frac{Q_0^2 A}{4^g} + \frac{{\rm Var}[q_g]}{4}\,,
\label{eq:recursive_var}
\end{equation}
where $Q_0^2 A$ is a constant. 
The first term goes rapidly to zero, meaning that eq.~\eqref{eq:recursive_var} spontaneously tends to the limit of constant splitting coefficient, or $A=0$, with fluctuations that are just rescaled from one generation to another, as described by eq.~\eqref{eqn:variance_with_qt}. 
At which generation this limit behaviour is reached depends on the values of the initial ${\rm Var}[q_1]$ and the constant $Q_0^2 A$. 
Note that the existence of such limit is yet another facet of the intrinsic property of our system that low-intensity flow rate fluctuations upstream are comparable to high-intensity fluctuations after a few bifurcations, which means that unsteady flow splitting is inevitably the more impactful the sooner it occurs. 

\begin{figure}
\centering
\includegraphics[width=0.75\textwidth]{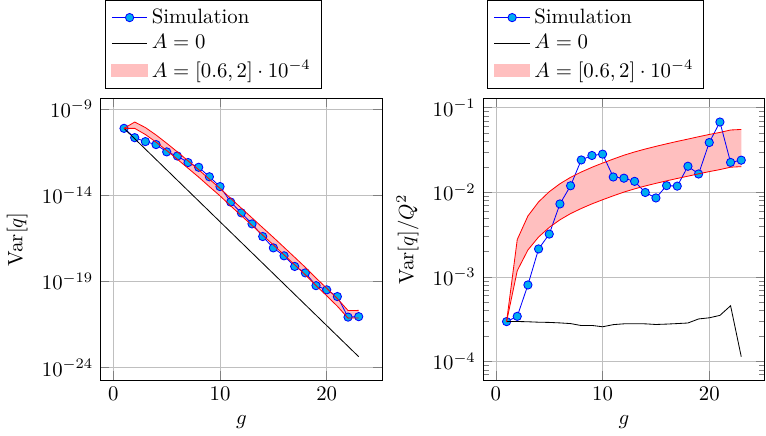}
\caption{Left: comparison of variance of the flow rate between simulation data and eq.~\eqref{eq:recursive_var}. Right: comparison of variance of the flow rate normalized with its mean value between simulation data and eq.~\eqref{eq:recursive_I}. The red areas denote estimates obtained with $A$ corresponding to standard deviation between $0.75\%$ and $1.25 \%$.}
\label{fig:varq}
\end{figure}

The same line of reasoning can be applied to the variance scaled with the mean, ${\rm Var}[q]/Q^2$, which is a measure of the relative intensity of fluctuations. 
The recursive formula for this quantity that is equivalent to eq.~\eqref{eq:recursive_var} is derived dividing eq.~\eqref{eqn:variance_with_qt_and_at} by $Q^2_{g+1}$ and using again $Q_g = Q_0/2^g$ and $S_g = 0.5$: 
\begin{equation}
	\begin{split}
	\frac{ {\rm Var}[q_{g+1}]}{Q^2_{g+1}} & = \frac{Q_g^2}{Q_{g+1}^2} A + \frac{S^2_g}{Q^2_{g+1}} {\rm Var}[q_g] \\
										  & = 4 A + \frac{{\rm Var}[q_g]}{4^{g+2} Q^2_0}	\,.
	\end{split}	
\label{eq:recursive_I}
\end{equation}
Contrary to the variance, the relative variance increases as long as $A>0$, illustrating how variable splitting coefficients are connected with larger fluctuations intensities. 
The variance and relative variance computed by our simulation are compared against the estimates provided by eq.~\eqref{eq:recursive_var} and eq.~\eqref{eq:recursive_I} in Fig.~\ref{fig:varq}. 
Values for the parameter $A$, which is a guess for the variance of splitting coefficients for all generations, are chosen to correspond to standard deviation between $0.75\%$ and $1.25 \%$; the particular case $A=0$, corresponding to constant $s$, is included as well.
The overall trend of the data is captured very well.
In particular, the trend of ${\rm Var}[q]$ eventually reaches the limit behaviour of $A=0$, but the initial amplification of flow rate fluctuations is such that the assumption of constant $s$ in the first generations causes a significant underestimate for both the variance and the scaled variance.
The details of the evolution along the main branch are however lost to the simplified models and, as illustrated in particular by the complex behaviour of the scaled variance, the covariance term appears to be necessary to obtain a better agreement with the simulation data. 

Further improvements of the models presented above would be of limited value, on consideration of the specific geometry of this test case.
Nonetheless, the results in Fig.~\ref{fig:varq} have the merit to show how a time-dependent splitting of the flow rate in the upper portion of the main branch creates amplified fluctuations of the local flow rate in all lower branches, and how these fluctuations cannot be easily attenuated, no matter how small the Reynolds number becomes. 
These models can also be employed to explore how changes of $S$ and ${\rm Var}[s]$ can affect fluctuations in later generations, providing in this way an indication of the robustness and generality of our results. 
This is analysis, which is left for \S~\ref{app:robustness}, confirms that the choice of $S\approx0.5$ in our case study has a relatively small effect on the general trends that we identified.

Here we can make one last observation on the implications of eq.~\ref{eqn:variance_with_qt_and_at} and the presence of the covariance term. 
This terms possibly introduces an additional reason for asymmetry in the lungs, which has been recently explained as the result of optimizing the filtering of dangerous particles in higher generations, at the expense of the overall efficiency in gas exchange~\cite{kundu-panchagnula-2023}. 
Introducing a non-zero correlation between flow rate and splitting makes little sense in a symmetric structure, where reducing fluctuation intensity in one branch implies increasing it in another, and inevitably in all following generations.
To the contrary, in a non symmetric tree, branches with a lower splitting coefficient can work as ``sponges'' for intense fluctuation in a given branch, if the proper correlation is introduced.

\subsection{Limitations of the current model and case study}
The question of whether the kind of velocity and flow-rate fluctuations that we observe are at least compatible with the physiological breathing cycle and the human anatomy immediately arises. 
These are actually two separate points: first of all, the statistically stationary flow rate of $Q_0=60 {\rm L/min}$ represents just one brief instance of inspiration, which is relevant for inhlation therapy but it is not reached in many subjects during tidal respiration; 
secondly, we must recognize that the lungs remain a closed system and, eventually, flow rates in all generations must be coupled with volume variations or, more precisely, with the rate of volume variations in the distal airways and alveolar sacs. 

On the first point, we can relay on dimensional analysis and the similarity principle to formulate a few observations. As long as chaotic flow is entering the bronchial tree, conditions at a given generation in our simulation are indicative of those found for rigid models at other statistically steady $Q_0$, with an appropriate shift of generation index.  
At a lower $Q_0$, the turbulent-to-laminar transition will be anticipated, and the initial level of disturbance in the bronchial tree will change. However, the initial separation between velocity, pressure, and flow-rate time scales, followed by their progressive coupling as the local Reynolds number decreases, is bound to occur as described in \S\ref{sec:res_history} and \S\ref{sec:res_scales}.
Likewise, if a chaotic regime is established in a portion of the model, time dependence in total resistance will still cause unsteady flow splitting, as described in \S\ref{sec:res_flow_fluctuations}, with fluctuation intensities rescaled according to the average values of each physical variable. 
We can see that a single variation of the average splitting, for instance that of $S_1$ from $0.5$ to $\approx 0.6$ or $\approx 0.4$, which would be closer to the average physiological values for the human respiratory system, will cause a similar rescaling as that caused by a change of $Q_0$.
A more intriguing question is whether our results apply for transient conditions. 
This case study was selected specifically to distinguish between the matters of scale separation and global time dependence, and reconciling them merits a dedicated effort. However, we can argue that the answer is to be found in the interplay between characteristic time scales identified in \S\ref{sec:res_scales} and the one(s) introduced with breathing.
In transient condition, we can define a time dependent bulk velocity $U^{bulk}(t)$, averaged over each generation. 
Under quasi-steady conditions, changes of $U^{bulk}(t)$ are much slower than convective time scales, and the flow evolution is still well represented by a succession of steady states at varying Reynolds numbers; in this case, the condition $T^{conv}/T^{diff}<1$ is still likely sufficient to identify where the Stokes regime is established. 
On the other hand, changes in the total flow rate are significant when $U^{bulk}(t)$ increases rapidly enough to avoid the diffusion of disturbances that would otherwise dissipate in a given generation and time interval, or, vice versa, when $U^{bulk}(t)$ decreases rapidly enough to cause the dissipation of disturbances that would have otherwise traveled through the generation. 
Note that, since convective time scales are local variables, whether or not acceleration and deceleration are immediately relevant for the state of flow depends on the generation as well as on breathing regimes: the case study examined here allowed to describe only the case of stationary transition. 
 
On the second point, it is fair to presume that the degree of fluctuations intensity in the distal generations in our simulation, while reasonably realistic for similar rigid models employed in experimental campaigns, may be at most an upper limit of fluctuations in the physiological case, in particular for the healthy lung. 
In a nutshell, the degree to which flow rate fluctuations with respect to the base-line breathing cycle can propagate is determined by the degree of coordination in structural deformation, as shown by the following example.
For the sake of simplicity, let us assume that dilatation occurs only at the level of single alveoli, and that it is governed by a perfectly sinusoidal rate of expansion with fixed amplitude and frequency but arbitrary phase shifts. 
That sinusoidal rate of expansion uniquely determines the flow rate downstream the bifurcation, where deformations become synchronous. 
Before that point, phase shifts in the expansion of a portion of alveoli would result in a more complex flow rate. 
Anisotropy will be introduced with more complex mechanisms than simple phase shifts in the lungs. 
Already in the healthy lung, the known tendency of alveoli to activate through an avalanche process~\cite{suki-etal-1994}, rather than with smooth and continuous deformation, is one candidate to generate flow rate fluctuations with much faster time scales than breathing. 
At the same time, the compression caused by gravity is known to introduce large-scale variations in the elastic propreties of the parenchima~\cite{bryan-1966}.  
More evident examples of multi-scale anisotropy are consequences of pathologies: for example, in the case of emphysema, the damage to a portion of the lung causes uneven stresses in the surrounding functional airways during respiration~\cite{suki-stamenovic-hubmayr-2011}.
Finally, even if it is conceivable that intermediate airways with compliance may elastically adjust to fast variation of flow rates, they also participate in the tissue distention, in mechanical equilibrium with the peripheral airways~\cite{pare-mitzer-2012}, thus changing both volume and resistance during breathing.
In this way, small asymmetries in the anatomy could be another source of unsteady flow partitioning between parallel branches.
Since none of these phenomena is explicitly taken into account, our results are not directly informative for the physiological lungs, for which complete high-fidelity simulations remain out of reach. 
However, they illustrate how flow regularization cannot entirely happen by virtue of the turbulent-to-laminar transition, because the laminar flow in deep generations fully retains the ability to transmit flow rate and pressure fluctuations.

\section{Conclusions}
\label{sec:conclusions}
A direct numerical simulation of the air flow in a rigid lung model including all its 23 generations, at a realistic steady inhalation flow rate of $60\,\rm L/min$, has been presented. It constitutes the first fully resolved simulation in space and time considering a range of scale representative of the entire bronchial tree, without relaying on a multi-scale approach, that would typically introduce artificial parameters, assumptions, and turbulence models.
We have studied in detail how the coexistence of various flow regimes in the bronchial tree is connected to the intensity of flow rate fluctuations at every generation.
The flow evolves from a turbulent-like regime, in the first generations, to an unsteady laminar regime and, eventually, to a Stokes flow in the deeper generations; these results are, broadly speaking, in agreement with the picture emerging from dimensional analysis and already accepted in the field. 
As the local Reynolds number diminishes, viscosity progressively suppresses small-scale velocity fluctuations until time dependence is expressed only as a simple pulsation of the laminar solution. 
We have also found that, perhaps unexpectedly, flow rate fluctuations retain a relatively broad range of active time scales and non-negligible intensity in the entire bronchial tree. 
These fluctuations ultimately arise from the unsteady flow partition caused by turbulence in higher bifurcations; contrary to velocity fluctuations, however, flow rate fluctuations survive through the branches, because of mass conservation.
This negates the conventional assumption that transition to a laminar regime implies suppression of all time scales but the one associated with breathing, which is adopted in many previous studies addressing deep lung generations.  

Beyond the simplifications of our model, the link between flow splitting and mass conservation is certainly relevant for simulations or {\em in-vitro} experiments that employ rigid models. 
Whether it carries the same importance for physiological bronchial trees is more difficult to assess.
On the one hand, there is a potential for fluctuations in flow splitting as long as there is a chaotic or turbulent-like regime in the first generations, which is typically the case in most breathing conditions.
On the other hand, the range of active scales that can be found in deep generations is inevitably interconnected and possibly constrained by tissue deformation.
It also has to be recognized that simulations under statistically steady conditions cannot capture the more complex scenarios that will arise under semi-periodic breathing conditions. 
A non-steady flow rate will cause a modification of the transition points between the various regimes, and guessing how this would interact with flow splitting is not trivial. 

Because of these additional effects that were not taken into account in our simulations, our results are not directly informative for the physiological lung. 
Nonetheless, the propagation of fast and intense flow rate fluctuations in deep generations found here highlights the risk of introducing assumptions on flow regimes hidden within turbulence modeling. 
The most straightforward example are multi-scale simulations, typically employing a RANS approach for the upper generations and a laminar flow assumption for the lower generations, with the two models matched through consistent average flow rate conditions. 
Even if the turbulence model was capable of guessing the exact turbulent kinetic energy, it would still, by design, provide an average velocity field and an average flow splitting. 
In intermediate generations, turbulent fluctuations are bound to dissipate, and the only surviving time scales would be those introduced with the choice of boundary conditions. 
Such an approach is then doomed to grossly underestimate active frequencies in lower generations, regardless of whether the correct flow regime is simulated in every generation. 
On the bright side, at least in principle it may also be easily improved. 
Assuming that a correlation between local turbulence intensity and flow splitting fluctuations is available, which can be measured via existing data or high-fidelity simulations limited to upper generations only, flow rate fluctuations can be included through the boundary conditions of laminar simulations, which remain perfectly suitable to describe the regime in intermediate generations.
A similar type of correction can be also included in stochastic models combining local time scales and fluctuation intensity to compute particle dynamics (discrete and continuous random walks) or to improve probability equations adopted in purely statistical deposition models.

\appendix
\section{Resolution assessment}
\label{app:resolution}

\subsection{Higher generations}
In this portion of the tree, the flow is mildly turbulent. 
A direct numerical simulation should discretize the equations on a grid fine enough to resolve all the dynamically relevant scales of motion.
Resolution requirements for turbulent duct flows are typically formulated by comparing the grid size with the so-called viscous length scale $\delta_\nu = \nu / u_\tau$, built with the fluid kinematic viscosity and the friction velocity $u_\tau$, derived from the wall-shear stress $\tau_w=\rho \nu ({\rm d} U/{\rm d} y)_{y=0}$ as $u_\tau=\sqrt{\tau_w/\rho}$ ~\cite{pope-2000} (in the expression above, $y$ should be intended as the wall-normal direction, and $y=0$ is the position of the wall).
For channel and pipe flows, plenty of information is available as to what resolution is enough to yield grid-independent statistics~\cite{pirozzoli-orlandi-2021}; generally speaking, an anisotropic grid size is often used, with a few viscous units in the wall-parallel direction and one viscous unit or slightly less in the wall-normal direction near the wall. In the present case, therefore, the geometric complexity suggests the use of an isotropic grid with $\Delta x=\Delta y=\Delta z$.

\begin{figure}[htp]
	\centering
	\includegraphics[width=0.55\textwidth]{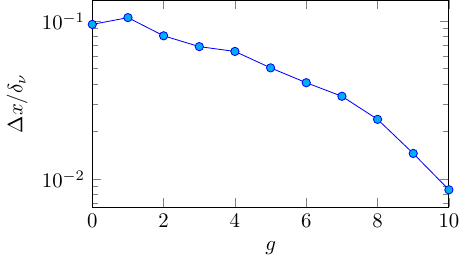} 
	\caption{Grid spacing scaled in viscous units. For $g=0$, $\Delta x / \delta_\nu $ is evaluated in the "trachea" region.}
	\label{fig:deltaxp}
\end{figure}

To provide a rough evaluation of the grid quality in our case, the first generations are considered as a sequence of turbulent homogeneous pipe flows, for which known correlations~\cite{dean-1978} between flow rate and wall friction can be used to estimate the friction velocity and the viscous length. As shown in Fig.~\ref{fig:deltaxp}, it turns out that, even at $g=1$, the grid spacing employed here is significantly finer than that required to resolve a classic turbulent pipe flow.

\subsection{Grid dependency at generations 21-23}
In deep generations, where the flow is laminar but unsteady, establishing resolution criteria is not trivial: the unsteadiness creates near-wall shear layers that can be very thin, challenging the adequacy of an isotropic mesh. Moreover, the diameter of the ducts shrinks, and the immersed-boundary algorithm is called to a significant work.

\begin{figure}[htp]
	\centering
	\includegraphics[width=0.55\textwidth]{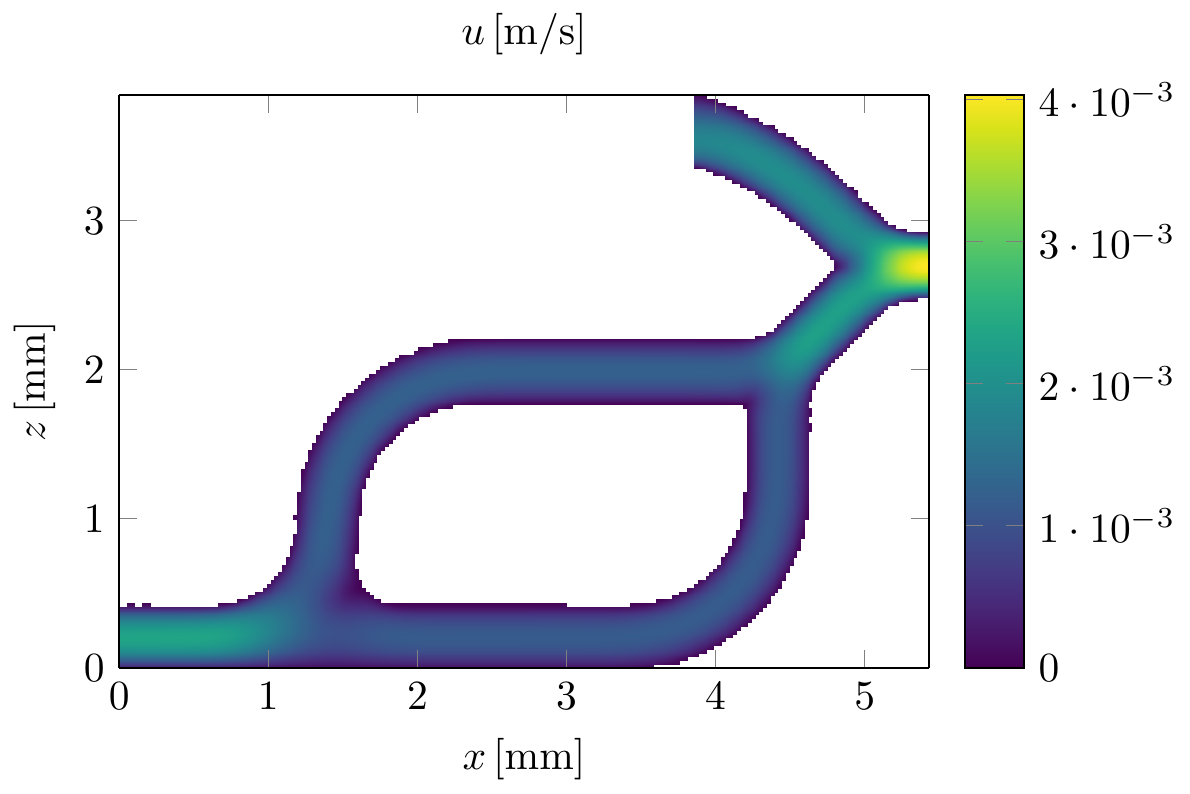} 
	\caption{Frontal section of the small model employed in the grid dependency analysis.}
	\label{fig:minimap}
\end{figure}

\begin{figure}[htp]
	\centering
	\includegraphics[width=0.66\textwidth]{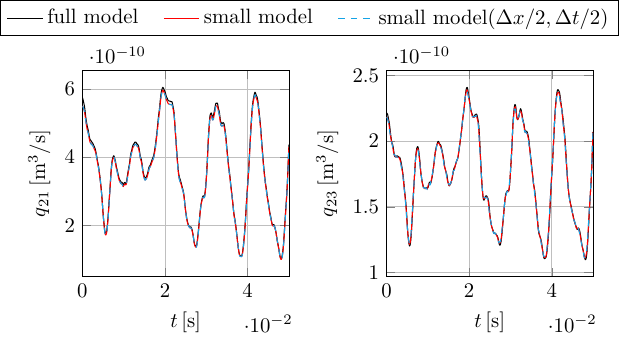} 
	\caption{Time history of $q_{21}$ and $q_{23}$ for the full model and the small model with two resolutions.}
	\label{fig:minires}
\end{figure}

To assess grid adequacy, the following grid-dependency test has been carried out on a small model including $g=22$ and $23$. 
The small model, illustrated in Fig.~\ref{fig:minimap}, is a cut of the full model, leaving only one inlet (the terminal section of $g=21$) and two outlets (the end of $g=23$ and the secondary branch of $g=22$).
The time history of the pressure signal, known at the inlet and the two outlets from the full-model simulation, is used as a driving boundary condition to set pressure values at the three patches of the small model, for which a conventional grid-independency study becomes possible.
A first simulation with the small model has been carried out with the same resolution used for the larger model, which amounts to $107 \times 9 \times 76$ grid points.
A fixed time step was used, equal to the average value of $\Delta t$ of the full simulation, even though using a time step derived from the CFL condition would allow a significantly larger $\Delta t$, owing to the much lower velocity in this portion of the domain. 
A second simulation with the small domain has been also carried out, using the same pressure history for the boundary conditions, but with twice the spatial and temporal resolution, thus increasing the computational size of the problem by more than one order of magnitude. 
The three flow rates across the patches of the small model can be compared against $q_{21}$, $q_{23}$, and $q_{22}^s$ from the full simulation. 
Note that such a direct comparison is possible because advection does not play a role in the Stokes regime, and the solution is uniquely determined by the pressure differences between patches, without memory of the flow history in previous generations. 
Fig.~\ref{fig:minires} shows the time history of $q_{21}$ and $q_{23}$; to simplify the visualization, only a brief time interval of $5 \cdot 10^{-2}\,{\rm s}$ is plotted, which is comparable to the active scales at this location. 
The time histories between the large model and the reduced model are virtually indistinguishable, attesting that the flow is indeed fully in the Stokes regime. Moreover, comparing the time history computed for the small model at the two resolutions confirms that even for the deep generations we are observing grid-independent results. 

\subsection{Estimate of turbulent viscosity}
Lastly, one can compute {\em a posteriori} the level of turbulent viscosity that would be introduced by a LES model, should the simulation become a LES. In the LES approach, the turbulent viscosity $\nu_t$ becomes zero when the grid is fine enough to achieve DNS resolution.
In practice, the LES model starts to work when $\nu_t >\nu$, and it often happens that $\nu_t \gg \nu$: the comparison between the turbulent viscosity and the molecular viscosity informs of the portion of the flow dynamics that is not captured naturally and requires modelling to balance insufficient resolution.
The turbulent viscosity $\nu_t$, defined as in the WALE sub-grid LES model~\cite{ducros-etal-1999}, has been computed from a number of flow fields. The maximum value of $\nu_t$ averaged through the fields has been found to be less than $1\%$ of the molecular viscosity in $g<3$, and even lower in the successive generations, while its maximum instantaneous value in the same region is $0.2 \nu$.

\section{Model for propagation of flow rate fluctuations}
\label{app:expansion}

\subsection{Validity of the first-order expansion}
\label{app:validity_expansion}

\begin{figure}[htp]
	\centering
	\includegraphics[width=0.45\textwidth]{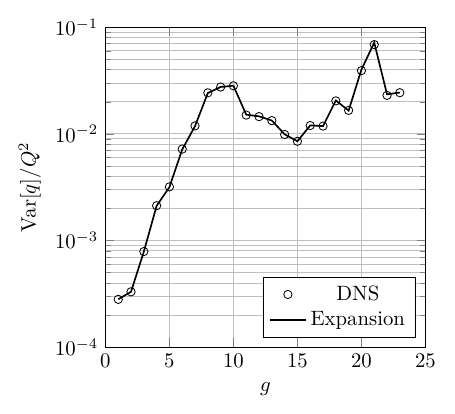} 
	\caption{Comparison between DNS data and prediction from eq.~\eqref{eq:expansion_I}.}
	\label{fig:expansion_vs_data}
\end{figure}

The model for propagation of flow rate fluctuations discussed in \S\ref{sec:res_flow_fluctuations} is derived from a first-order expansion of the flow rate and splitting coefficients. 
We can now compare against DNS data the intensity of flow rate fluctuations defined from eq.~\eqref{eqn:variance_with_qt_and_at}, i.e.: 
\begin{equation}
	 \frac{{\rm Var}[q_{g+1}]}{Q_{g+1}^2} \approx \frac{Q_g^2 {\rm Var}[s_{g}] + S_g^2 {\rm Var}[q_{g}] + 2 S_g Q_g {\rm Cov}[s_g, q_g]}{Q_{g+1}^2} \,,
	 \label{eq:expansion_I}
\end{equation}
to assess the validity of this expansion, before introducing any other assumption or approximation (in this expression, $Q_{g+1}\approx Q_{g}S_g$).
Without the additional simplifications that eventually led to eq.~\eqref{eq:recursive_I}, eq.~\eqref{eq:expansion_I} predicts fluctuation intensity in generation $g+1$ from data from variance and covariance of flow rate and splitting in generation $g$.
The comparison is shown in Fig.~\ref{fig:expansion_vs_data}. 
Note that errors in this comparison do not propagate, because ${\rm Var}[q]$ from DNS data is used at each step to predict ${\rm Var}[q_{g+1}]$, rather than ${\rm Var}[q]$ predicted from the previous step, so that we can assess the expansion validity for each generation.
Predictions from expansion exhibit excellent agreement with simulation data, the largest discrepancy being found at the very end of the bronchial tree, which is inevitably a sensitive location for the extreme low flow rate.

\subsection{Sensitivity to local splitting average and variance}
\label{app:robustness}
The excellent agreement between flow rate variance prediction from the first order expansion suggests that the models derived in \S\ref{sec:res_flow_fluctuations} can account for the most relevant contributions to flow splitting. 
Thus, they can be used to investigate the sensitivity of our results w.r.t.\ the choice of average splitting coefficients -- which are an outcome of the calibration procedure -- and of splitting variances -- which are a property of the geometry. 

\begin{figure}[htp]
	\centering
	\includegraphics[width=0.55\textwidth]{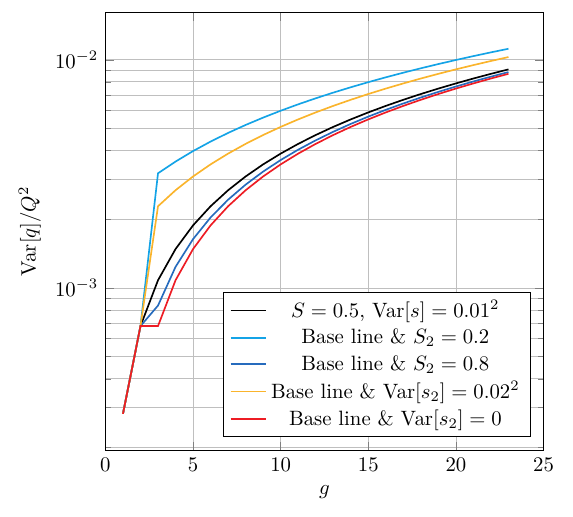} 
	\caption{Comparison between predictions from the recursive formula~\eqref{eq:recursive_I} with $S=0.5$ and $ {\rm Var}[s]=0.01^2$, and altering splitting average or fluctuations at $g=2$.}
	\label{fig:sensitivity}
\end{figure}

To this end, we start comparing the prediction from the recursive formula~\eqref{eq:recursive_I} with uniform splitting average $S=0.5$ and splitting variance $ {\rm Var}[s]=0.01^2$ with those obtained with modification of either $S_2$ or $ {\rm Var}[s_2]$, which is shown in Fig.~\ref{fig:sensitivity}. 
The covariance term is always set to $0$ for simplicity.
Local changes of $S$ or ${\rm Var}[s]$ inevitably have repercussion deeper in the tree. 
In fact, $S_2<0.5$ implies that a lower portion of the flow remains in the main branch, corresponding to an increase of the ratio ${\rm Var}[q_2]/Q_2^2$ in particular and of ${\rm Var}[q]/Q^2$ in general, with respect to the base line model. 
On the contrary, $S_2>0.5$ implies that a higher portion of the total flow passes through the main branch with, keeping everything else unchanged, a corresponding decrease of ${\rm Var}[q]/Q^2$.
Increasing or decreasing ${\rm Var}[s_2]$ has the same effect, respectively, of decreasing or increasing $S_2$, with the additional note that we can in principle set ${\rm Var}[s_2]=0$, which corresponds to get a constant intensity through generation $2$. 
It is also worth stressing that modifying $S$ in a single-branch model may be misleading in that $S$ for two branches must add to $1$, so in a realistic tree branches with some $S<0.5$ cannot be introduced without also having as many $S>0.5$ in parallel branches, whereas ${\rm Var}[s]$ is the same for the two daughter branches.  

\begin{figure}[htp]
	\centering
	\includegraphics[width=0.55\textwidth]{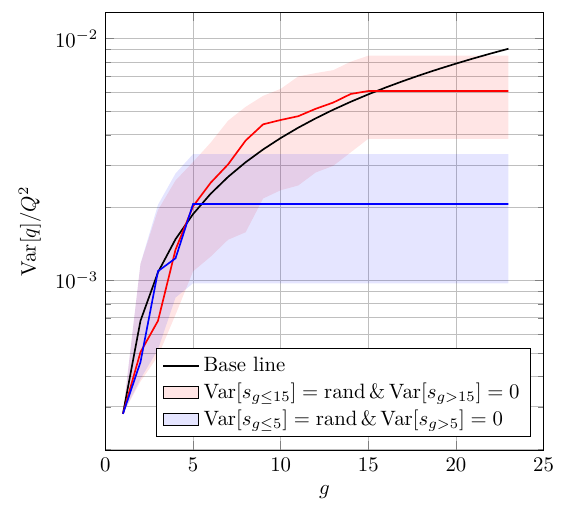} 
	\caption{Comparison between predictions from the recursive formula~\eqref{eq:recursive_I} with $S=0.5$ and $ {\rm Var}[s]=0.01^2$, and when $ {\rm Var} [s]$ is randomized in the interval between $0.005^2$ and $0.015^2$ and set to zero after $g=5$ or $g=15$. Solid lines denote selected cases, while the colored regions indicate the envelopes of $100$ cases.}
	\label{fig:envelops}
\end{figure}

The trend identified with uniform $S$ and ${\rm Var}[s]$ will always appear somewhat artificially regular when compared against the DNS data.  
Fig.~\ref{fig:envelops} shows a comparison between predictions with the conventional $S=0.5$ and $ {\rm Var}[s]=0.01^2$ and two ensembles of $100$ randomized trees. 
Without significant loss of generality for this tests, average splitting is still kept uniform at $S=0.5$. 
The variance is randomized for each generation, within the interval corresponding to $0.5\%$ and $1.5\%$ splitting standard deviation. 
The two ensembles differ in that one has ${\rm Var}[s]=0$ for all bifurcation below $g=15$ and the other has ${\rm Var}[s]=0$ for all bifurcation below $g=5$.
This last modification, which would correspond to a perfectly synchronous deformation in that portion of airways, is included here because it is the only alteration that leads to a significantly different trend from that observed in the DNS data. 

The analysis presented here was kept at a qualitative level because the specific selection of $S$ and ${\rm Var}[s]$ are a relatively minor assumption, compared to the statistically steady state total flow rate and rigid airways.
Nonetheless, it confirms that $S$ and ${\rm Var}[s]$ are not themselves critical parameters that bear an excessive influence on the results we presented.

\section*{Acknowledgments}
The authors thank M. Reddemann and H. Schmitz (Institute of Heat and Mass Transfer, RWTH Aachen University) for helpful discussions during the preparation of the manuscript. 
This research has been partially supported by the Italian Research Center on High Performance Computing Big Data and Quantum Computing (ICSC), project funded by European Union - NextGenerationEU - and National Recovery and Resilience Plan (NRRP) - Mission 4 Component 2. Computational time was also provided by the ICSC, through the International Foundation Big Data and Artificial Intelligence for Human Development (IFAB), and by the CINECA Italian Supercomputing Center through the ISCRA project StocLung.


\section*{Ethical statement}
The authors have nothing to report.

\section*{Conflict of Interest}
The authors declare that they have no known competing financial interests or personal relationships that could have appeared to influence the work reported in this paper.

\section*{Funding Information}
This research has been partially supported by the Italian Research Center on High Performance Computing Big Data and Quantum Computing (ICSC), project funded by European Union - NextGenerationEU - and National Recovery and Resilience Plan (NRRP) - Mission 4 Component 2. 

\section*{Data Availability Statement}
The data that support the findings of this study are available from the corresponding author upon reasonable request.

\section*{Author Contribution Statement}
\textbf{M. Atzori:} conceptualization, data curation, formal analysis, investigation, methodology, software, validation, visualization, writing - original draft, writing - review \& editing.
\textbf{E. Gallorini:} methodology, software, validation, writing - original draft, writing - review \& editing.
\textbf{C. Cottini:} funding acquisition, project administration, resources, supervision, writing - review \& editing.
\textbf{A. Benassi:} conceptualization, funding acquisition, investigation, methodology, project administration, resources, writing - review \& editing.
\textbf{M. Quadrio:} conceptualization, funding acquisition, investigation, methodology, project administration, resources, software, supervision, writing - original draft, writing - review \& editing.

\bibliography{Lungs.bib,Wallturb.bib}

\section*{Short Bio of the Authors} 
\par \textbf{Marco Atzori} obtained a master’s degree in physics at the University of Genoa (2017) and a doctorate in engineering mechanics at the Royal Institute of Technology -- KTH (2021). He is now a researcher at Politecnico di Milano.
\par \textbf{Emanuele Gallorini} obtained a master’s degree (2019) and a PhD (2024) in aerospace engineering at Politecnico di Milano. He is now a postdoctoral fellow at the Laboratory Mechanical Fluid De Lille. 
\par \textbf{Ciro Cottini} obtained a master’s degree in physics at the University of Padova (2000) and joined Chiesi Group in 2010, where he is now head of the Digital Data \& Modelling department. He obtained a PhD in 2020 at the University of Bologna. 
\par \textbf{Andrea Benassi} obtained a master’s degree (2005) and a doctorate in physics at the University of Modena and Reggio Emilia (2008). He held researcher positions in academic institutions and joined R\&D of Chiesi Group in 2015. 
\par \textbf{Maurizio Quadrio} obtained a PhD in Aerospace Engineering in 1993. He became full professor of fluid mechanics at Politecnico di Milano in 2015, where he coordinates the activities of the Instability and Flow Control Lab.

\end{document}